\documentclass[
aps,
prb,
twocolumn,
amsmath,
amssymb
]{revtex4-2}

\usepackage{graphicx}
\usepackage{dcolumn}
\usepackage{subcaption}
\usepackage{bm}

\begin{document}

\title{\textbf{Machine Learning of Topological Insulator and Anderson Insulator in One-Dimensional Extended Su-Schrieffer-Heeger Chain} 
}%

\author{Zhekai Yin}
\affiliation{Department of Physics, Xiamen University Malaysia, 43900 Sepang, Selangor, Malaysia}

\author{C. K. Ong}
\email{ckong@xmu.edu.my}
\affiliation{Department of Physics, Xiamen University Malaysia, 43900 Sepang, Selangor, Malaysia}
\affiliation{Key Laboratory for Magnetism and Magnetic Materials of the Ministry of Education, Lanzhou University, Lanzhou 730000, China}

\date{\today}

\begin{abstract}
We study disorder effects in the extended Su–Schrieffer–Heeger (SSH) model using a convolutional neural network (CNN) trained on reduced correlation matrices (RCMs) of disorder-free systems to predict winding number phase diagrams in systems with off-diagonal and diagonal disorder. The trained CNN model generalizes to chiral-symmetry-preserving off-diagonal disorder system but fails in the presence of chiral-symmetry-breaking diagonal disorder system. Using principal component analysis (PCA) of the RCM feature space, we demonstrate that disorder-free and symmetry-preserving systems share overlapping feature manifolds, whereas symmetry-breaking disorder causes them to diverge. Inverse participation ratio (IPR) and energy spectrum analysis further demonstrate that off-diagonal disorder preserves topological edge states, whereas diagonal disorder drives a transition to an Anderson insulator. Our results show that the OOD behavior of a CNN trained on clean systems can be understood through the evolution of the RCM feature space under symmetry-preserving and symmetry-breaking disorder, with IPR and energy-spectrum analyses providing the corresponding physical interpretation.
\end{abstract}

\maketitle


\section{Introduction}

Topological insulators, characterized by topological invariants and robust boundary states\cite{hasan2010colloquium,qi2011topological}, and Anderson insulators, characterized by bulk state localization\cite{anderson1958absence,abrahams1979scaling}, have reshaped our understanding of quantum materials\cite{li2009topological}. The Su–Schrieffer–Heeger (SSH) model provides a paradigmatic one-dimensional example, where localized edge states are protected by chiral symmetry and classified by an integer winding number\cite{li2014topological,meier2016observation,asboth2016schrieffer}. The SSH can be extended to include diagonal disorder and off-diagonal disorder to examine effects of various perturbations. In the off-diagonal disorder perturbation, the chiral symmetry of the linear chain remains intact and exhibits topological phase transition. In the diagonal disorder perturbation, the chiral symmetry of the linear chain is broken and bulk localizations appear, exhibiting Anderson insulator transition.  

In this paper,  we train a supervised convolutional neural network (CNN) on analytic reduced correlation matrices (RCMs) of disorder-free SSH models with long-range hopping as input data set, together with  three winding numbers ($\nu \in \{0,1,2\}$) as targets. The trained CNN model is then generalized to study out-of-distribution (OOD) systems, successfully predicting the phase diagram for off-diagonal disorder system while failing to do so for diagonal disorder system. Previous studies have established that neural networks can learn topological invariants in clean chiral systems and some have further investigated out-of-distribution (OOD) generalization in disordered SSH-type models. These studies provide important foundations for using neural networks to learn topological information and to analyze their reliability under distribution shifts. Zhang et al.\cite{zhang2018machine} demonstrated that neural networks can learn global winding-number patterns from local momentum-space Hamiltonian data in clean one-dimensional chiral-symmetric systems. Their work mainly addresses the learnability of the winding-number formula itself, rather than the robustness of learned representations under disorder-induced distribution shifts. Cybinski et al.\cite{cybinski2025characterizing} studied OOD generalization in the disordered SSH model using eigenstate-intensity inputs, with emphasis on the reliability and interpretability of CNN predictions under disorder. Their work analyzes why some CNNs learn disorder-robust features while others rely on less transferable correlations. In contrast, the present work considers an extended SSH chain with long-range hopping, which supports three winding-number sectors \(\nu=0,1,2\), and uses reduced correlation matrices as real-space, entanglement-related inputs. More importantly, we explicitly compare chiral-symmetry-preserving off-diagonal disorder with chiral-symmetry-breaking diagonal disorder within the same framework. Our focus is therefore not only on whether the CNN can classify topological phases, but also on how the RCM feature space evolves when the symmetry protecting the topological phase is preserved or broken. In this sense, our results are consistent with previous demonstrations that neural networks can capture topological information, while extending the analysis to a multi-winding, RCM-based, disorder-dependent setting.\\
Besides, in order to understand the role of chiral symmetry playing in space feature extraction in CNN technique, we perform principal component analysis (PCA) of the RCM feature space, which reveals overlapping manifolds for disorder-free and symmetry-preserving disorder systems, but exhibits a clear separation when symmetry is broken. \\
To investigate why the CNN fails to generalize to OOD cases where symmetry is broken, we note that PCA merely classifies the similarity of feature distributions rather than their physical nature. We, therefore, proceed to study the localization properties of single-particle eigenstates via inverse participation (IPR) and the energy spectrum across the hopping parameters. It shows that edge states survive only in the presence of chiral symmetry, while diagonal disorder breaks the chiral symmetry and drives a transition to an Anderson insulator.
\section{Su-Schrieffer-Heeger Model}
\subsection{Hamiltonian}
\begin{figure}[tbp]
    \centering
    \includegraphics[width=0.8\linewidth]{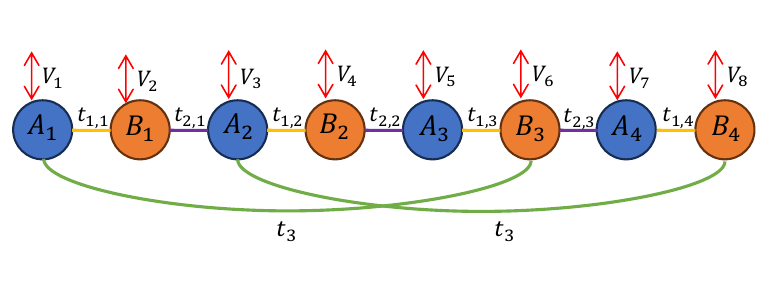}
    \caption{Schematic diagram of SSH model with long range hopping. $A_i,B_i$ are sites of the $i$-th unit cell. $\{t_{1,1},t_{1,2},t_{1,3},...\}$ are the intra-cell hopping amplitudes, $\{t_{2,1},t_{2,2},t_{2,3},...\}$ are the inter-cell hopping amplitudes, $t_3$ is the long range hopping amplitude, $\{V_1,V_2,V_3,...\}$ are the on-site potentials.}
    \label{SSH}
\end{figure}

We consider an extended Su–Schrieffer–Heeger (SSH) model on a one-dimensional chain with 
$N$ unit cells (total $2N$ sites) under open boundary conditions (OBC). The Hamiltonian in real space is:
\begin{equation}
\begin{aligned}
\hat{H} = & \sum_{n=1}^{N} \left(t_{1,n} \hat{c}_{n,A}^{\dagger} \hat{c}_{n,B} + \mathrm{h.c.}\right)  + \sum_{n=1}^{N-1} \left(t_{2,n} \hat{c}_{n,B}^{\dagger} \hat{c}_{n+1,A} + \mathrm{h.c.}\right) \\
& + t_{3} \sum_{n=1}^{N-2} \left(\hat{c}_{n,A}^{\dagger} \hat{c}_{n+2,B} + \mathrm{h.c.}\right)  + \sum_{i=1}^{2N} V_i \hat{c}_i^{\dagger} \hat{c}_i,
\end{aligned}
\end{equation}
where $\hat{c}_{n,\alpha}$ ($\alpha=A,B$) creates a spinless fermion on sublattice 
$\alpha$ of the $n$-th unit cell. The first two terms describe the standard dimerized hopping, where $t_{1,n}$ is the intra-cell amplitude and $t_{2,n}$ is the inter-cell amplitude. We introduce a long range hopping $t_3$ between $A_n$ and 
$B_{n+2}$, which extends the model beyond the conventional binary topological classification. $V_i$ describes the on-site potential at each site $i$, which accounts for external electric fields or structural inhomogeneities in the 1D chain. Disorder is incorporated via random modulations:
\begin{equation}
    t_{1,n}=t_1+W_1\omega_n^{(1)}, \quad t_{2,n}=t_2+W_2\omega_n^{(2)}, \quad V_i=V_{\text{on}}\omega_i^{(3)},
\end{equation}
where the variables $\omega$ are independent and identically distributed (i.i.d.) according to a uniform distribution on the interval $[-0.5, 0.5]$. The parameters $W_1,W_2$ control off-diagonal disorder strength, while $V_{\text{on}}$ introduces diagonal (on-site) disorder strength. The structure of this long range hopping SSH model is sketched in Fig.~\ref{SSH}.

\subsection{Winding Number}\label{nu}
In the disorder-free model ($W_1=W_2=V_{\text{on}}=0$), the system is translationally invariant. Under periodic boundary conditions (PBC), the Bloch Hamiltonian is:
\begin{equation}\label{eq3}
H(k) = \begin{pmatrix}
0 & h^*(k) \\
h(k) & 0
\end{pmatrix},
\quad
h(k) = t_1 + t_2 e^{-ik} + t_3 e^{-i2k}.
\end{equation}
The system belongs to the chiral (AIII) symmetry class \cite{shapiro2022continuum,han2020topological} and is characterized by the integer winding number:
\begin{equation}\label{eq4}
\nu = \frac{1}{2\pi i} \oint_{\mathrm{BZ}} dk\, \partial_k \ln h(k),
\end{equation}
which counts the number of times the complex function $h(k)$ winds around the origin as $k$ traverses the Brillouin zone. According to Cauchy Argument Principle\cite{ahlfors1979complex,chen2020elementary,lee2022winding}, the winding of a complex function around the origin is related to its zeros and poles inside a closed contour. Specifically, the integrand $\partial_k \ln h(k)$ in Eq. \eqref{eq4} maps the unit circle in the $z$-plane ($z=e^{ik}$) to the winding of the characteristic polynomial $t_3z^2+t_2z+t_1$ in the complex $h$-plane. Hence, the winding number $\nu$ is fully determined by the number of roots of the polynomial $t_3z^2+t_2z+t_1=0$ that lie inside the unit circle $(|z|<1)$ . This yields a rich phase diagram with three distinct topological sectors $(\nu \in \{0,1,2\})$, in contrast to the standard SSH model ($t_3=0$) which only supports $\nu=0$ or 1.

In the disorder-free model, the winding number $\nu$ is rigorously defined under PBC via the momentum-space Bloch Hamiltonian (Eq.~\eqref{eq4}). However, experimental probes and numerical simulations of real materials are typically performed under OBC, where the system exhibits edge states and lacks translational symmetry. Therefore, it doesn't have a well-defined Brillouin zone.
However, following the bulk–boundary correspondence, the number of zero-energy edge states under OBC is strictly determined by the bulk winding number $\nu$ calculated under PBC\cite{asboth2016short}. In our machine learning training, we therefore use PBC-computed $\nu$ as label together with OBC-derived reduced correlation matrix (RCM) as the input data set.

\subsection{Symmetry and Stability Under Disorder}
The topological classification in Eq.~\eqref{eq4} relies crucially on chiral symmetry, which requires the Hamiltonian to be off-diagonal in a suitable basis. This symmetry is preserved when only off-diagonal disorder is present ($V_{\text{on}}=0$). In this case, the winding number remains well-defined, and the system exhibits a symmetry-protected topological phase robust against weak disorder \cite{cayssol2013floquet,prodan2011disordered,bernevig2013topological}.\\
By contrast, diagonal disorder ($V_{\text{on}}\neq0$) explicitly breaks chiral symmetry. Consequently, the bulk-boundary correspondence breaks down, and the topological phase diagram collapses into a featureless insulating regime dominated by Anderson localization \cite{hatsugai1993edge,max2019bulk,anderson1958absence,lee1985disordered}.\\
This difference between robustness under symmetry-preserving disorder and fragility under symmetry-breaking disorder will serve as the central physical principle to interpret the out-of-distribution (OOD) generalization behavior of our machine learning model in Sec.~\ref{sec4}. Specifically, OOD generalization\cite{cybinski2025characterizing,zhang2018machine} refers to the CNN’s ability to correctly identify topological phases when tested on disordered configurations that are entirely absent from the disorder-free training set. 
\section{Convolutional Neural Network (CNN)}

We employ a supervised 2D CNN to classify the topological phases of the system. The model uses RCMs as the input data, mapping it to winding number $\nu \in \{0, 1, 2\}$. The primary aim is to evaluate the model's OOD generalization.

\subsection{Input Data: Reduced Correlation Matrix (RCM)}\label{RDM}

We employ the single-particle correlation matrix (RCM) as the input data for our CNN. For a non-interacting system at half-filling, the many-body ground state $|\Psi_0\rangle$ is a Slater determinant fully characterized by the correlation matrix $C_{ij} = \langle \Psi_0 | \hat{c}^\dagger_i \hat{c}_j | \Psi_0 \rangle$. In practice, $C$ is constructed as a projection operator onto the subspace of the $N$ occupied single-particle eigenstates:\begin{equation}C = \sum_{n=1}^{N} |\psi_n\rangle \langle \psi_n|,\end{equation}where $\{|\psi_n\rangle\}$ are the $N$ lowest-energy eigenstates obtained from the diagonalization of the single-particle Hamiltonian. To reduce the input dimensionality while preserving topological signatures, we define our input data $C_{\text{sub}} \in \mathbb{R}^{N \times N}$ as the sub-block of $C$ restricted to sublattice $A$.\\
As the input for our CNN, we utilize the RCM $C_{\text{sub}}$, which serves as a computationally efficient representation of the many-body reduced density matrix (RDM)\cite{yin2025calculation,peschel2003calculation,sharma2008reduced}. Because our system consists of non-interacting fermions, $C_{\text{sub}}$ contains the complete information of the RDM via the single-particle entanglement spectrum. Thus, by training on $C_{\text{sub}}$, the network is effectively learning the entanglement spectrum $\{\xi_k\}$ that characterize the topological phases\cite{peschel2003calculation,peschel2009reduced,chung2000density,crampe2020entanglement}.

\subsection{CNN Architecture}
\begin{figure}[tbp]
    \centering
    \includegraphics[width=0.5\textwidth]{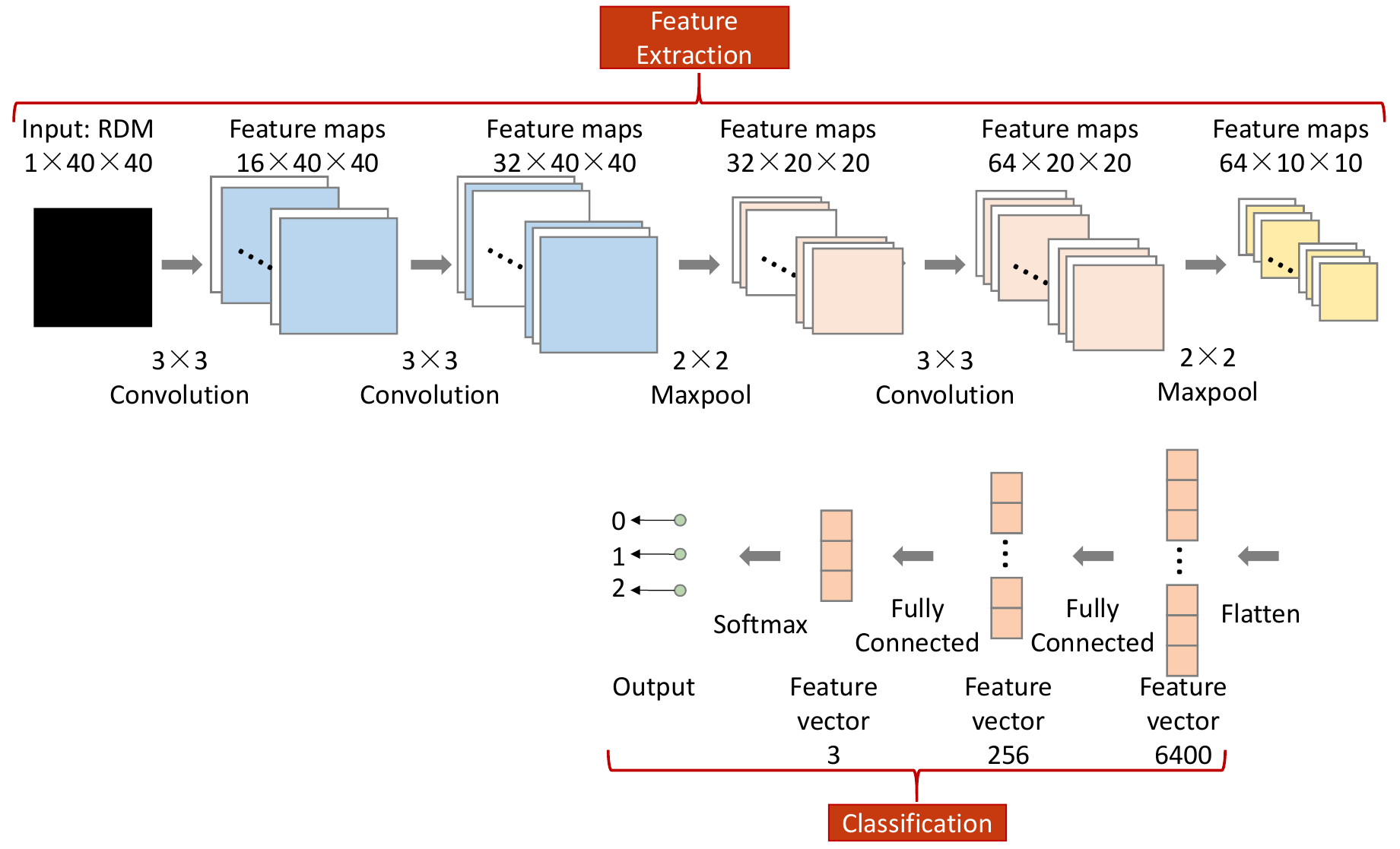}
    \caption{Schematic diagram of CNN architecture. The input is RCM of size $1 \times 40 \times 40$. The feature extraction stage consists of three convolutional layers (using $3 \times 3$ kernels) and two max-pooling layers (using $2 \times 2$ kernels) to progressively extract high-level topological features while reducing spatial dimensionality. The final feature maps of size $64 \times 10 \times 10$ are flattened into a 6400-dimensional feature vector. This vector is passed through two fully connected layers, reducing the dimensionality to 256 and finally to 3. A Softmax activation function is applied at the output layer to provide the predicted probabilities for winding numbers $\{0, 1, 2\}$.}
    \label{fake}
\end{figure}

Our classifier is a standard 2D CNN\cite{o2015introduction,wu2017introduction,bouvrie2006notes}, illustrated in Fig.~\ref{fake}. The input to the network is the $N \times N$ RCM ($N=40$), treated as a single-channel grayscale image and reshaped into a tensor of size $1 \times N \times N$. The architecture comprises three convolutional blocks: each block contains a $3 \times 3$ convolutional layer with zero padding, followed by a ReLU activation; max-pooling with kernel size $2 \times 2$ is applied after the second and third blocks, progressively reducing the spatial dimensions from $40 \to 20 \to 10$. The resulting feature maps (of size $64 \times 10 \times 10$) are flattened into a 6400-dimensional vector and passed through two fully connected layers (6400 $\to$ 256 $\to$ 3) with ReLU activation after the first fully connected layer. The final layer outputs three numerical numbers for the three winding number classes $\nu = 0, 1, 2$. These are the raw numbers the model calculates for each winding number before they are converted into actual probabilities using softmax function.

The model is trained exclusively on the disorder-free system data set using the Adam optimizer and cross-entropy loss:
\begin{equation}
    \mathcal{L} = -\frac{1}{|\mathcal{S}|} \sum_{(x,y) \in \mathcal{S}} \sum_{k=0}^{2} \mathbb{I}(y = k) \log p_k(x),
    \label{eq:cross_entropy}
\end{equation}
where $\mathcal{S}$ denotes the input dataset, $x$ is the RCMs, $y \in \{0,1,2\}$ is the true winding number label, $p_k(x)$ is the predicted probability for class $k$ (obtained via softmax), and $\mathbb{I}(\cdot)$ is the indicator function. Specifically, $\mathbb{I}(y=k)$ yields 1 if the true winding number $y$ matches the class index $k$, and 0 otherwise. Training employs early stopping when the loss falls below $10^{-5}$.\\
To examine the robustness with respect to neural-network initialization, we repeat the training procedure for \(N_{\rm seed}=10\) independent random seeds. For each seed, the CNN architecture and training hyperparameters are kept the same, while the network parameters are independently initialized.

\section{CNN Predictions under Disorder}\label{sec4}

\begin{figure}[tbp]
    \centering
    \includegraphics[width=0.35\linewidth]{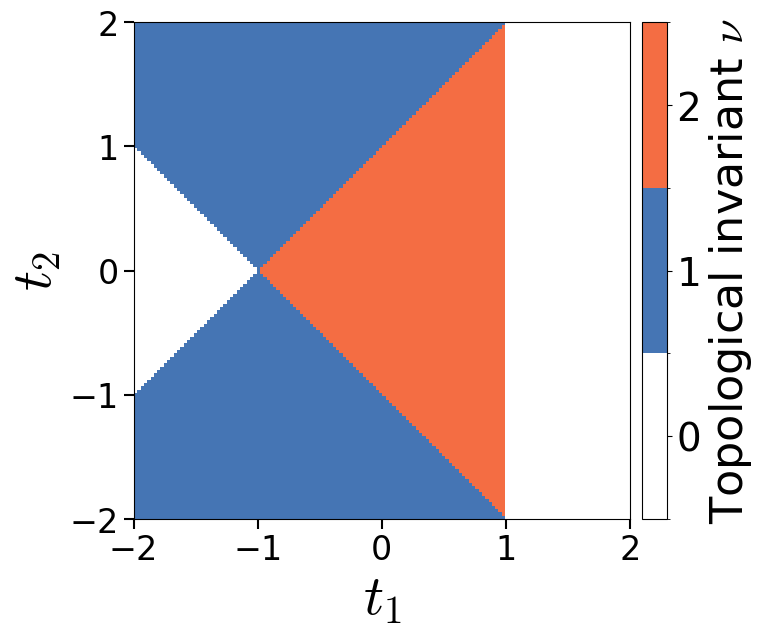}
    \caption{Analytical winding number map of disorder-free system}
    \label{exact}
\end{figure}

The training data set is constructed from the disorder-free SSH model, where the Hamiltonian parameters are uniformly sampled within the range $t_1, t_2 \in [-2, 2]$. For each parameter point $(t_1, t_2)$, we compute the RCM as described in Sec.~\ref{RDM}. And each RCM is labeled with its corresponding winding number $\nu$, calculated by the method in Sec.~\ref{nu}.\\
The winding number phase diagram for this disorder-free SSH model is presented in Fig.~\ref{exact}.

To evaluate the generalization ability of the trained CNN, we subject the model to both off-diagonal and diagonal disorder regimes. For each disordered configuration, we compute the corresponding RCM and feed it as input into the network. The CNN then maps it to winding number. By scanning the $(t_1, t_2)$ parameter space under varying disorder strengths, we construct predicted phase diagrams of the winding number $\nu$.\\
Crucially, for each parameter point \((t_1,t_2)\), we evaluate \(K=50\) independent disorder realizations for each of the \(N_{\rm seed}=10\) independently trained CNNs. The softmax probabilities are averaged over both disorder realizations and network seeds,
\begin{equation}
    \bar{p}_{\nu}(t_1,t_2)
=
\frac{1}{N_{\rm seed}K}
\sum_{s=1}^{N_{\rm seed}}
\sum_{r=1}^{K}
p_{\nu}^{(s,r)}(t_1,t_2),
\end{equation}
where \(p_{\nu}^{(s,r)}\) is the softmax probability for winding-number class \(\nu\) obtained from the \(s\)-th independently trained CNN and the \(r\)-th disorder realization. The predicted winding number is then assigned as
\begin{equation}
    \nu_{\rm pred}(t_1,t_2)
=
\arg\max_{\nu\in\{0,1,2\}}
\bar{p}_{\nu}(t_1,t_2).
\end{equation}
The confidence and normalized entropy are computed from the same averaged probability distribution,
\begin{equation}
    C(t_1,t_2)=\max_{\nu}\bar{p}_{\nu}(t_1,t_2),
\end{equation}
and
\begin{equation}
    H_{\rm norm}(t_1,t_2)
=
-\frac{1}{\log 3}
\sum_{\nu=0}^{2}
\bar{p}_{\nu}(t_1,t_2)\log \bar{p}_{\nu}(t_1,t_2).
\end{equation}
The average confidence quantifies the robustness of topological signatures in the RCM against disorder: high values ($>0.9$) indicate that edge-state correlations remain dominant across disorder realizations, while low confidence signals proximity to phase boundaries or symmetry-breaking regimes where topological features become ambiguous. Conversely, the normalized entropy measures topological mixing induced by disorder—peaking when individual realizations fluctuate between distinct winding numbers (e.g., residual edge modes competing with bulk localization), thereby serving as an indicator of symmetry-breaking transitions where chiral protection fails.\\
The robustness of these seed-averaged predictions with respect to independent CNN initializations is further quantified in Appendix~\ref{app:seed_robustness}, where we report the mean confidence, mean normalized entropy, and their seed-to-seed fluctuations for different disorder strengths.

\subsection{OOD Prediction for Off-Diagonal Disorder Case}
We first apply off-diagonal disorder, which preserves chiral symmetry. We set different values for off-diagonal disorder strength $(W_{\text{off}}=W_1=W_2=0.1,1.0,2.0,3.0,4.0,5.0)$, while $V_{\text{on}}=0$. Under this condition, the winding number  remains a well-defined topological invariant, and the phase diagram is expected to persist, with critical regions broadened by disorder-induced fluctuations.

\begin{figure}[tbp]
    \centering
    
    \begin{subfigure}[t]{0.15\textwidth}
        \centering
        \includegraphics[width=0.9\textwidth]{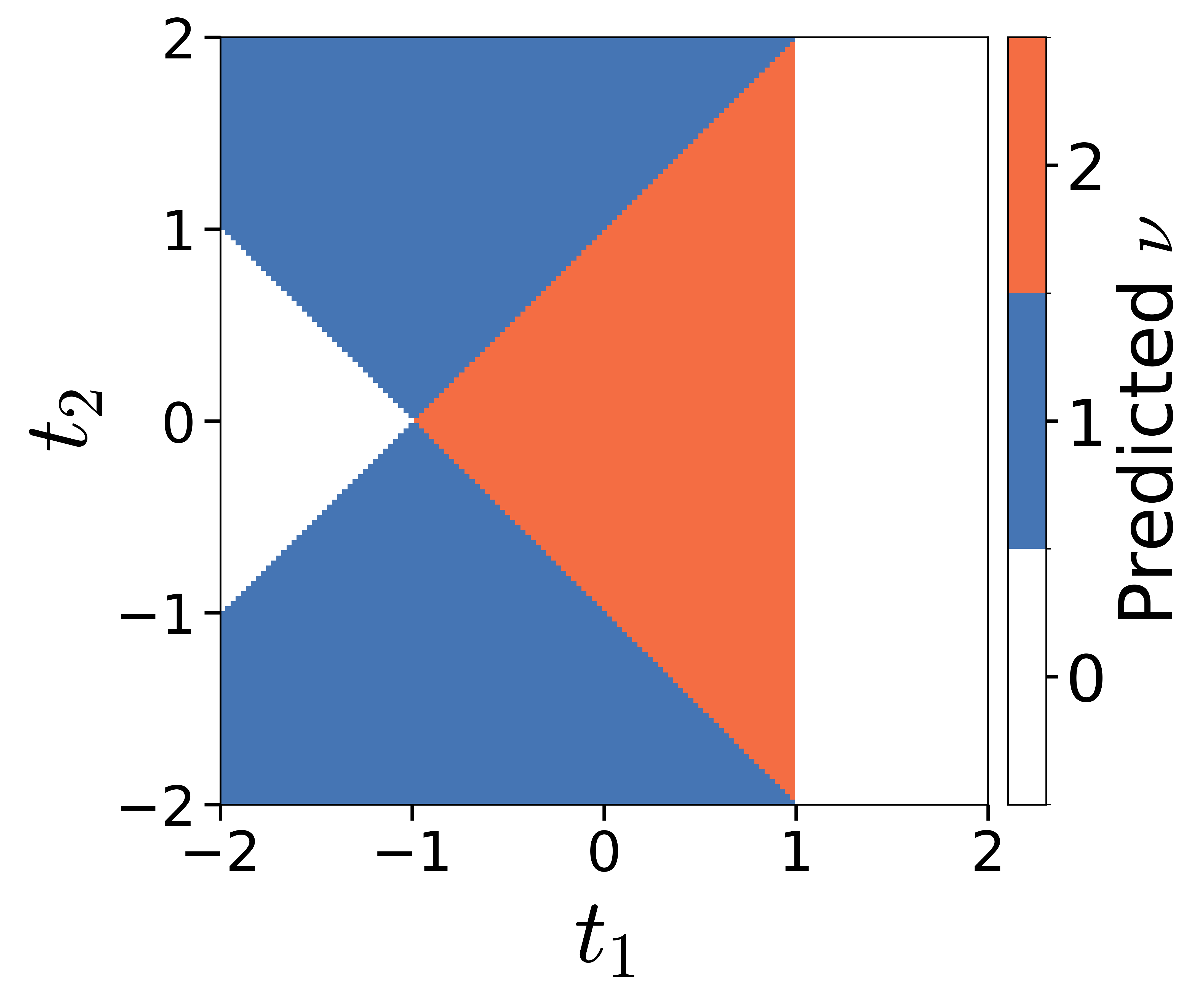}
        \caption{$\nu$ map for $W_{\text{off}}=0.1$}
        \label{fig:1a}
    \end{subfigure}
    \hfill
    \begin{subfigure}[t]{0.15\textwidth}
        \centering
        \includegraphics[width=0.9\textwidth]{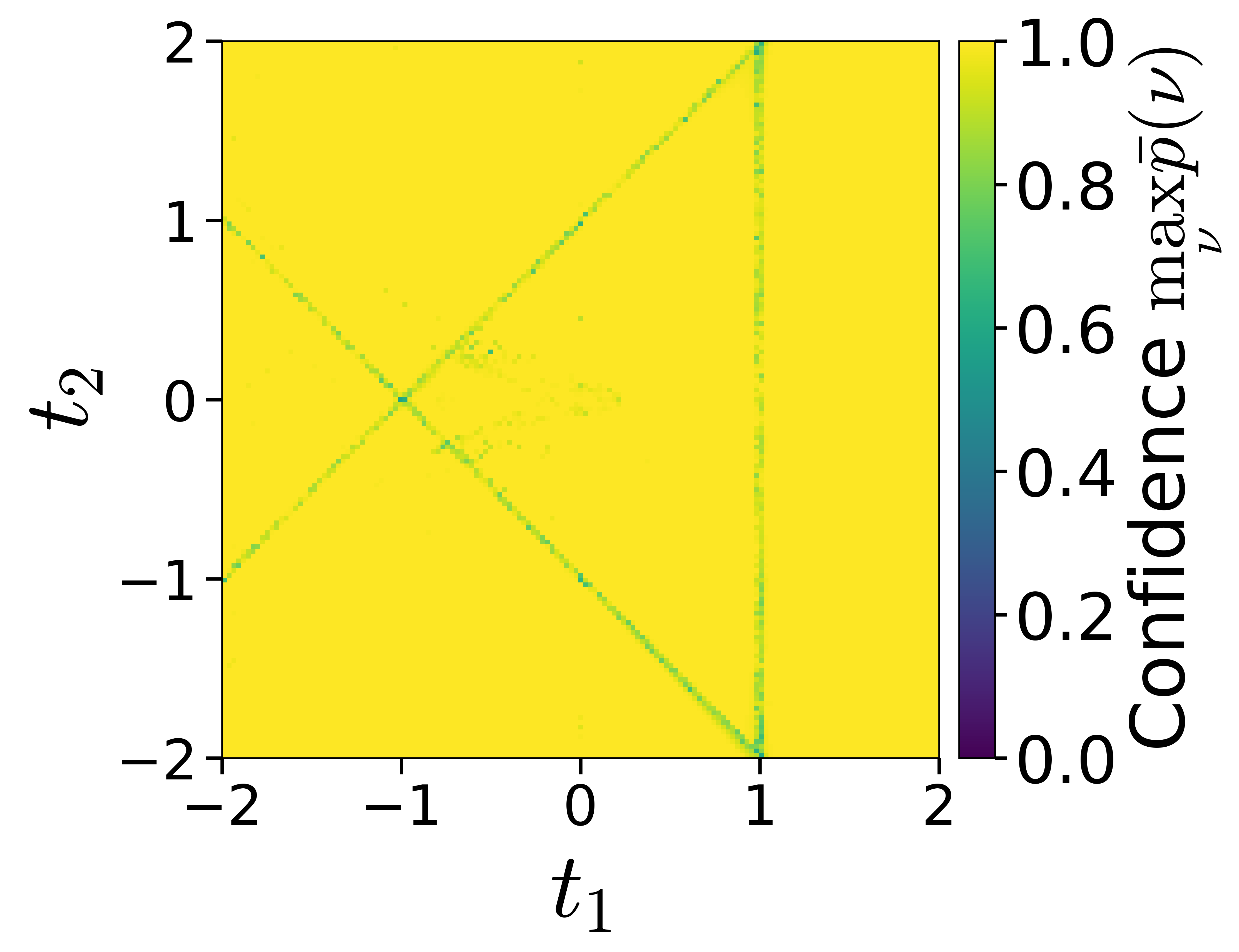}
        \caption{Confidence map}
        \label{fig:1b}
    \end{subfigure}
    \hfill
    \begin{subfigure}[t]{0.15\textwidth}
        \centering
        \includegraphics[width=0.9\textwidth]{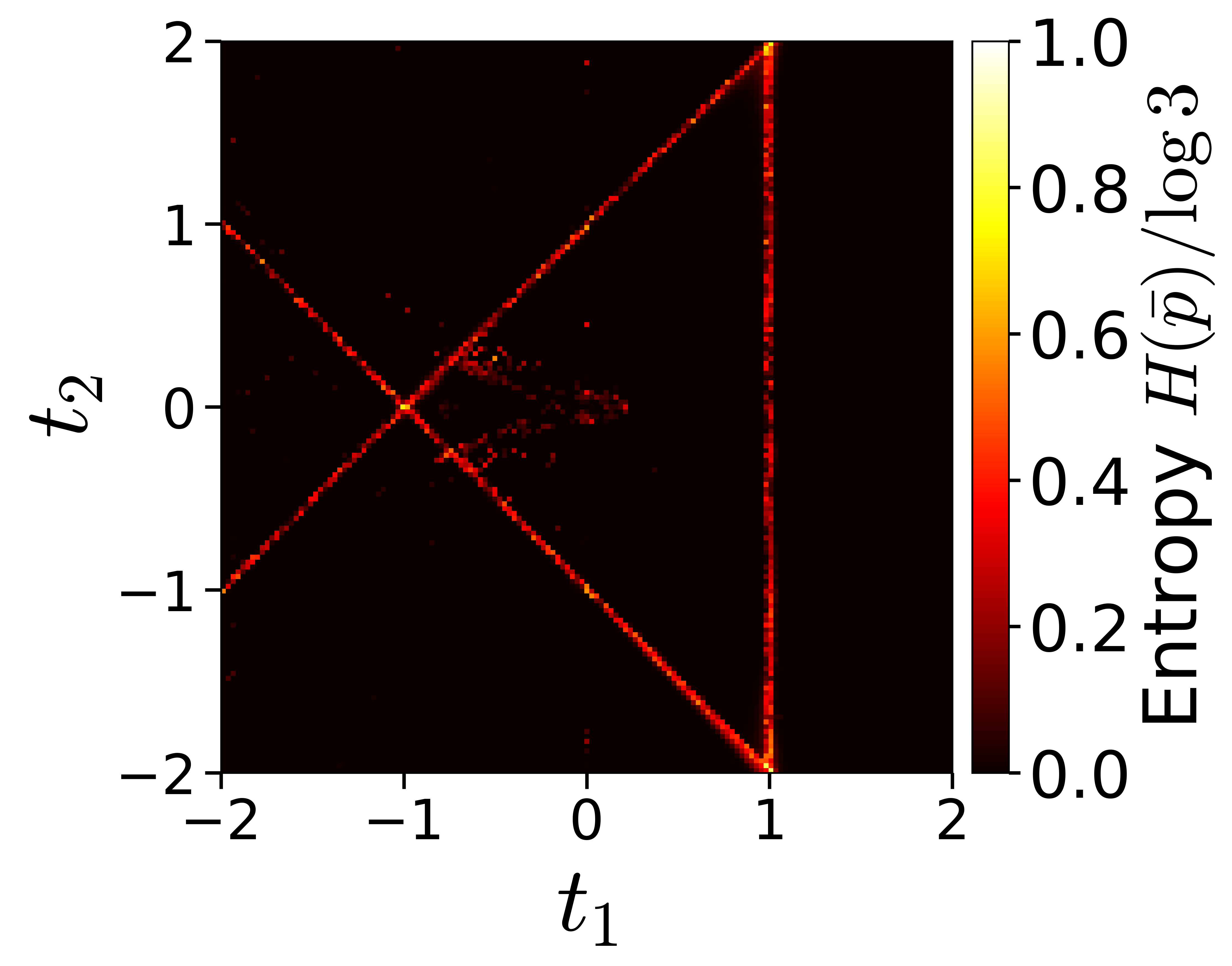}
        \caption{Entropy map}
        \label{fig:1c}
    \end{subfigure}
    
    \caption{Visualization results for $W_{\text{off}}=0.1$: (a) winding number $\nu$ map, boundary is the same as disorder-free system, (b) confidence map indicating prediction certainty, confidence is very high throughout the map, (c) entropy map representing prediction uncertainty, entropy is very low throughout the map. All maps are obtained by averaging the softmax probabilities over \(N_{\rm seed}=10\) independently initialized CNNs and \(K=50\) disorder realizations for each parameter point.}
    \label{fig:row1_results}
\end{figure}

As shown in Fig.~\ref{fig:1a} for $W_{\text{off}}=0.1$, the CNN’s predicted phase map remains strikingly similar to the disorder-free case. All three topological sectors are clearly identifiable, and phase boundaries align closely with the analytical disorder-free critical lines. The confidence (Fig.~\ref{fig:1b}) is high deep within each phase and drops only near boundaries, while the entropy (Fig.~\ref{fig:1c}) peaks precisely in those transition regions. This behavior originates from the closing of the topological gap at criticality: at phase boundaries the bulk gap vanishes, causing edge states to hybridize with the continuum and rendering RCM features ambiguous between competing winding numbers. Deep within a phase, by contrast, the finite topological gap protects edge-state signatures in the RCM, ensuring consistent classification across disorder realizations.\\
As $W_{\text{off}}$ increases (see Appendix \ref{App2}), the boundaries gradually blur, but the topological structure remains accurate until $W_{\text{off}}\approx3.0$.

This demonstrates that the CNN naturally generalizes to symmetry-preserving disorder. It does not merely memorize disorder-free features; it has learned a representation that is physically robust because the underlying topological classification itself is robust. 

\subsection{OOD Prediction for Diagonal Disorder Case}
\begin{figure}[htbp]
    \centering
    
    \begin{subfigure}[t]{0.15\textwidth}
        \centering
        \includegraphics[width=0.9\textwidth]{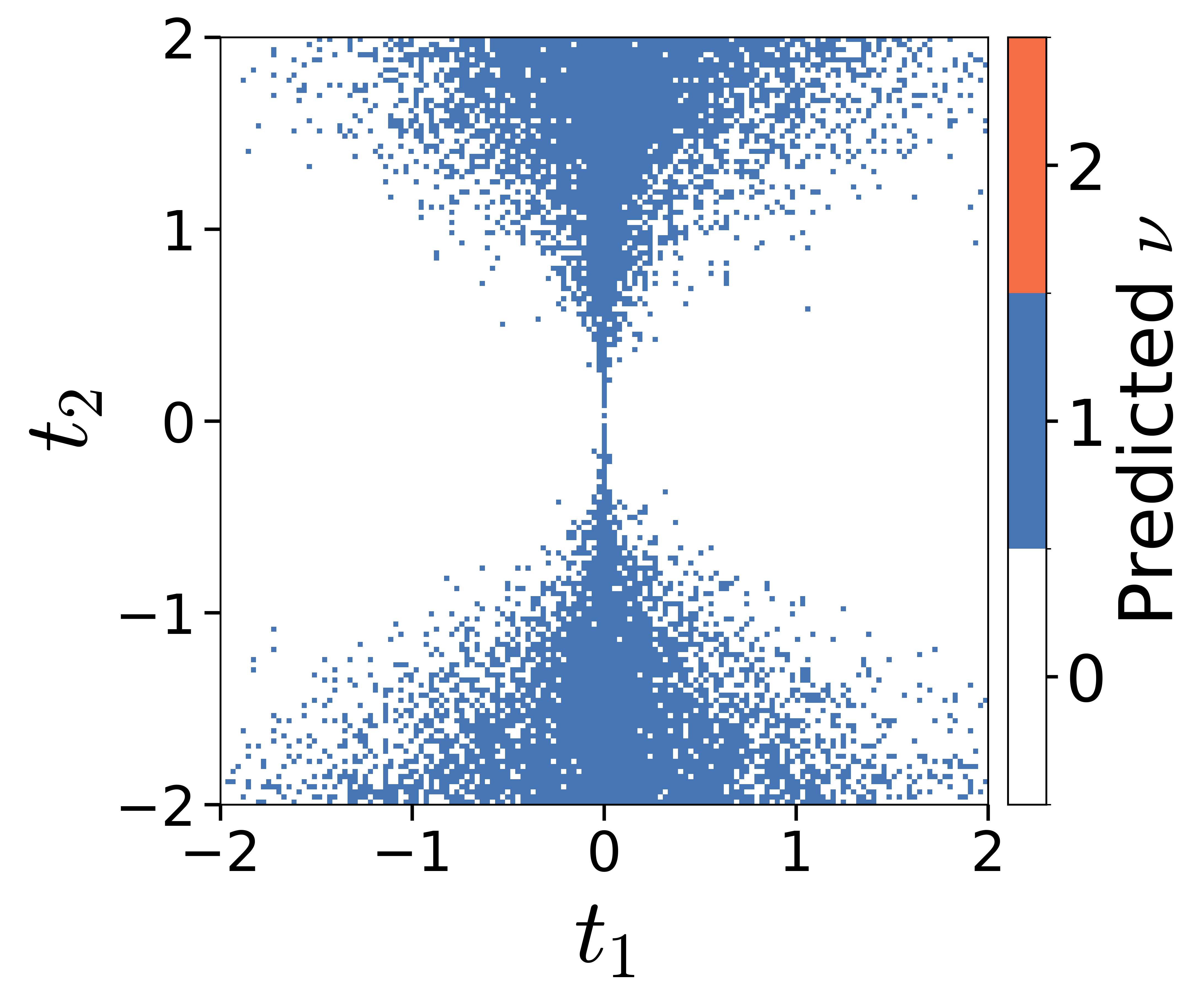}
        \caption{$\nu$ map for $V_{\text{on}}=0.1$}
        \label{4a}
    \end{subfigure}
    \hfill
    \begin{subfigure}[t]{0.15\textwidth}
        \centering
        \includegraphics[width=0.9\textwidth]{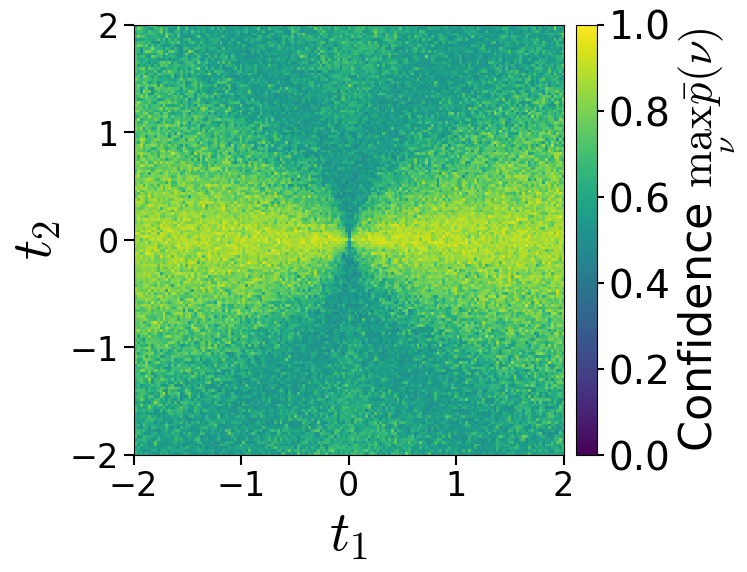}
        \caption{Confidence map}
        \label{4b}
    \end{subfigure}
    \hfill
    \begin{subfigure}[t]{0.15\textwidth}
        \centering
        \includegraphics[width=0.9\textwidth]{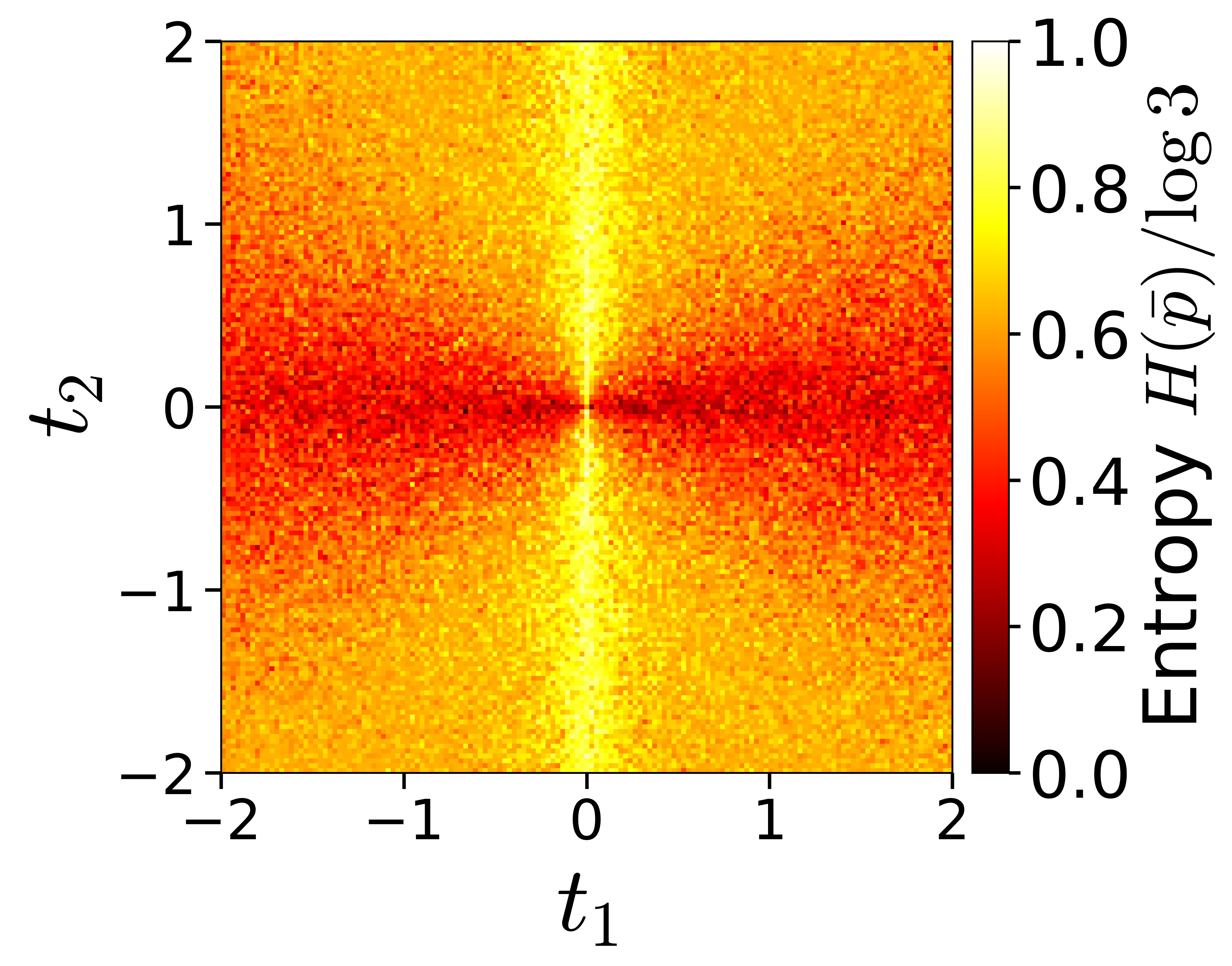}
        \caption{Entropy map}
        \label{4c}
    \end{subfigure}
    \caption{Visualization results for $V_{\text{on}}=0.1$: (a) winding number $\nu$ map, the clean-system phase structure is no longer reliably reproduced and fragmented residual regions appear near the original phase boundaries, (b) confidence map indicating prediction certainty, showing reduced and spatially inhomogeneous prediction confidence, (c) entropy map representing prediction uncertainty, showing enhanced uncertainty over a broad region of the parameter space. All maps are obtained by averaging the softmax probabilities over \(N_{\rm seed}=10\) independently initialized CNNs and \(K=50\) disorder realizations for each parameter point.}
\end{figure}

Next, we introduce diagonal disorder \(V_{\rm on}=0.1\), while keeping \(W_{\rm off}=0\). In contrast to off-diagonal disorder, diagonal disorder breaks chiral symmetry. As shown in Fig.~\ref{4a}, the clean-system topological phase structure is no longer reliably reproduced. The predicted phase map becomes dominated by a single winding-number class, with fragmented residual regions near the original phase boundaries rather than well-defined topological sectors. The confidence is reduced and spatially inhomogeneous (Fig.~\ref{4b}), while the normalized entropy is enhanced over a broad region of the parameter space (Fig.~\ref{4c}). These results indicate that the clean-trained CNN becomes substantially less reliable once the disorder breaks the symmetry protecting the topological phases.\\
This failure is not a flaw of the CNN—it is physically correct. In the absence of chiral symmetry, the winding number $\nu$ is no longer a conserved quantity\cite{kitaev2009periodic,prodan2011disordered,bernevig2013topological}; the bulk-boundary correspondence breaks down, and the system transitions into an Anderson insulator, which will be discussed in Sec.~\ref{sec6}. There is no true topological phase structure to predict. The reduced confidence and enhanced entropy therefore indicate that the RCM features learned from the clean chiral-symmetric system are no longer compatible with the symmetry-broken disorder ensemble.

\subsection{OOD Prediction for Mixed Off-diagonal and Diagonal Disorder}
\begin{figure}[htbp]
    \centering
    
    \begin{subfigure}[t]{0.15\textwidth}
        \centering
        \includegraphics[width=0.9\textwidth]{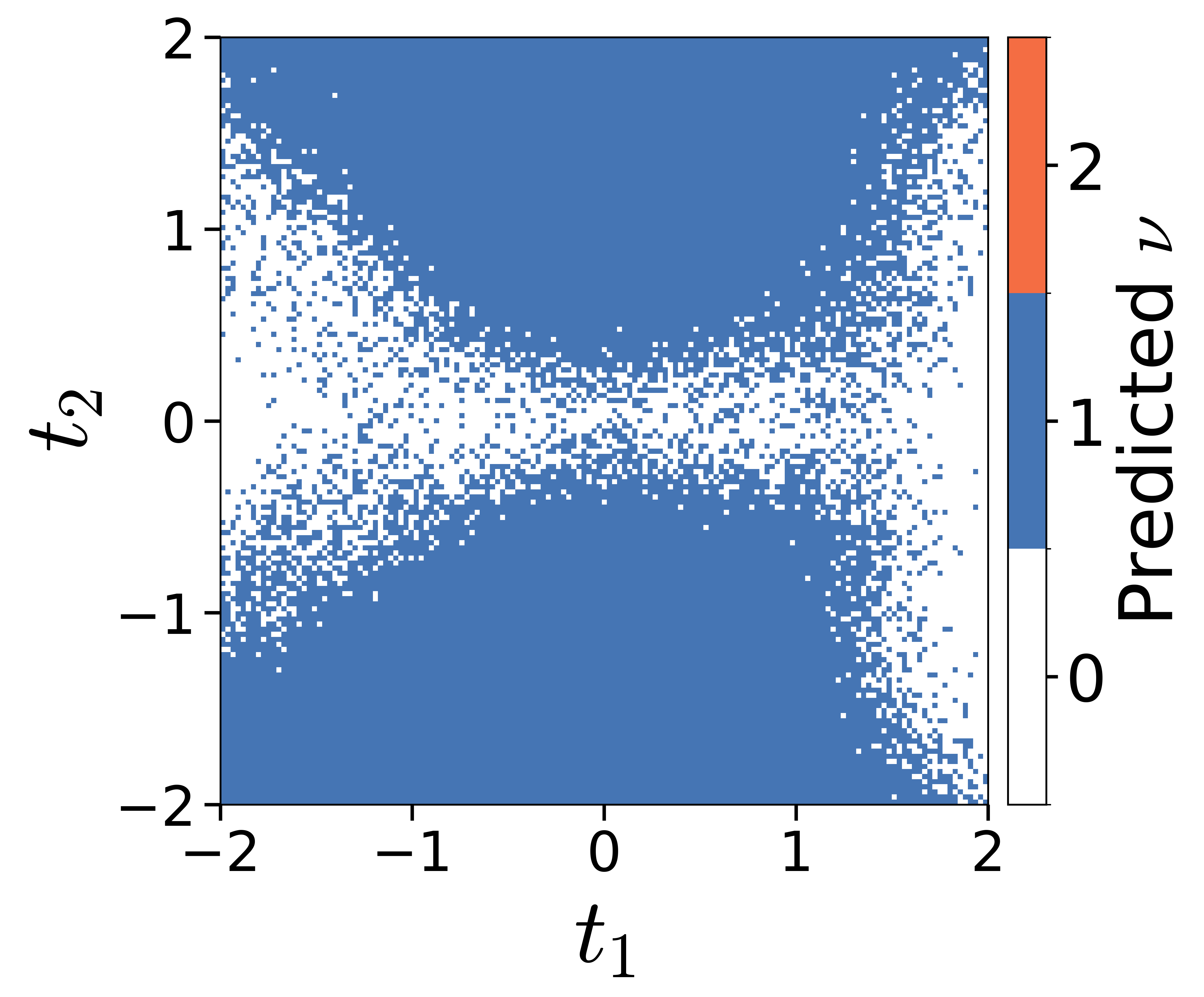}
        \caption{$\nu$ map for $V_{\text{on}}=0.01$}
        \label{mixed1}
    \end{subfigure}
    \hfill
    \begin{subfigure}[t]{0.15\textwidth}
        \centering
        \includegraphics[width=0.9\textwidth]{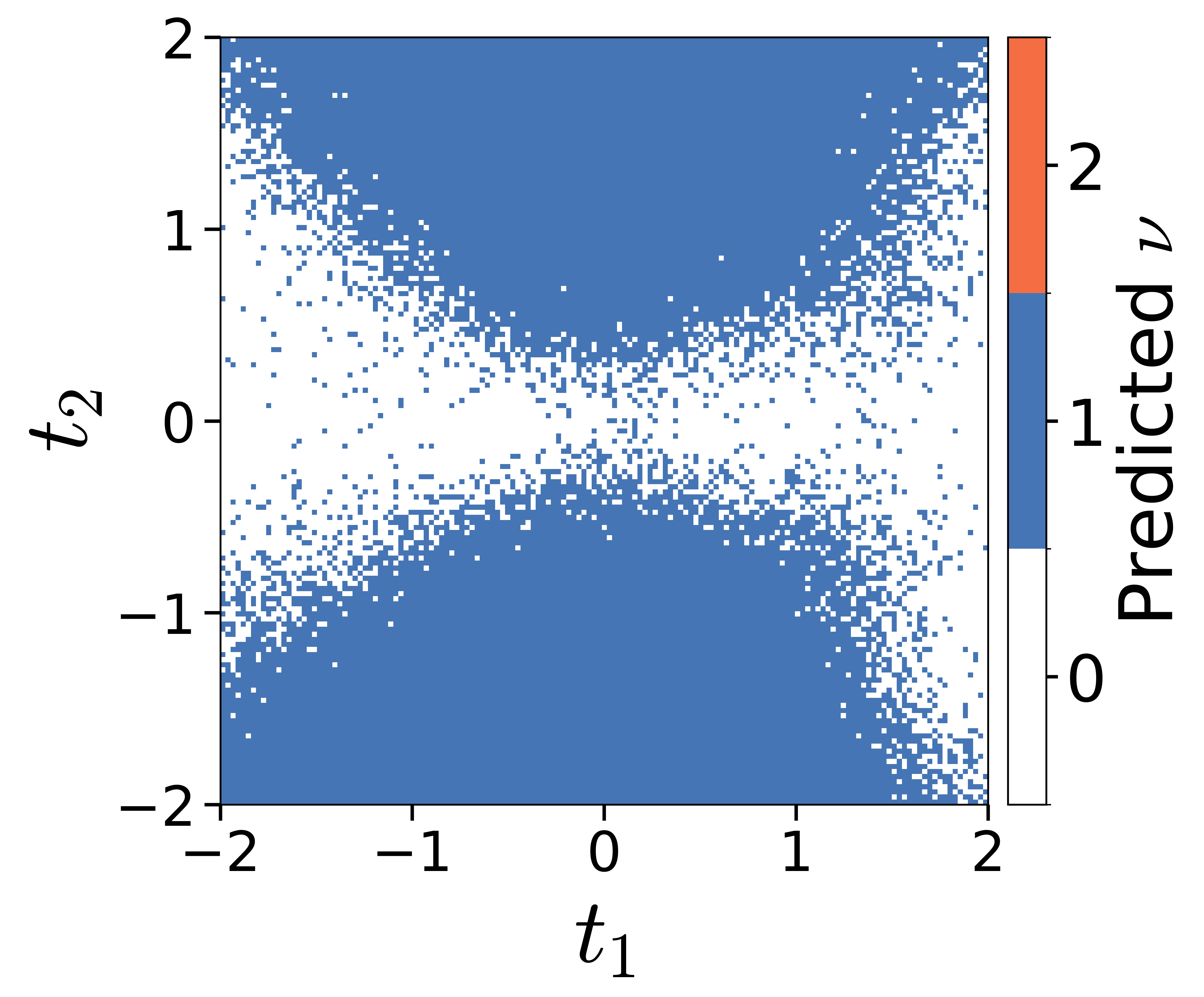}
        \caption{$\nu$ map for $V_{\text{on}}=0.1$}
        \label{mixed2}
    \end{subfigure}
    \hfill
    \begin{subfigure}[t]{0.15\textwidth}
        \centering
        \includegraphics[width=0.9\textwidth]{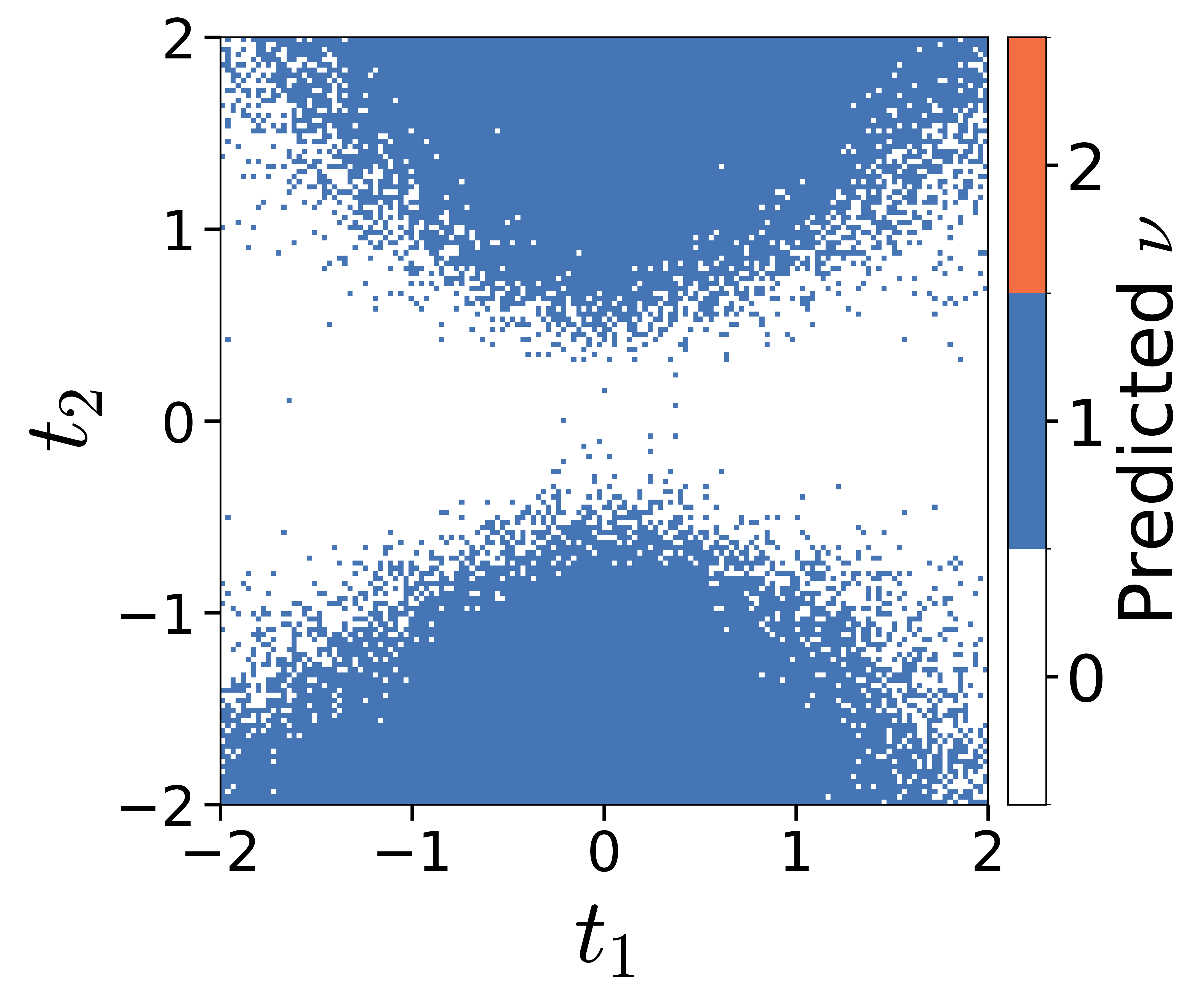}
        \caption{$\nu$ map for $V_{\text{on}}=1.0$}
        \label{mixed3}
    \end{subfigure}
    \caption{Winding number $\nu$ map for mixed-disorder at fixed $W_{\text{off}}=1.0$ and varying $V_{\text{on}}$: (a) $V_{\text{on}}=0.01$, (b) $V_{\text{on}}=0.1$, (c) $V_{\text{on}}=1.0$. The three maps share a similar overall pattern dominated by fragmented \(\nu=1\) regions, but increasing \(V_{\rm on}\) progressively blurs the residual sector boundaries and enhances spatial fragmentation. All maps are obtained by averaging the softmax probabilities over \(N_{\rm seed}=10\) independently initialized CNNs and \(K=50\) disorder realizations for each parameter point.}
    \label{fig:mixed_phase_maps}
\end{figure}

\begin{figure}[htbp]
    \centering
    \begin{subfigure}[t]{0.22\textwidth}
        \centering
        \includegraphics[width=\linewidth]{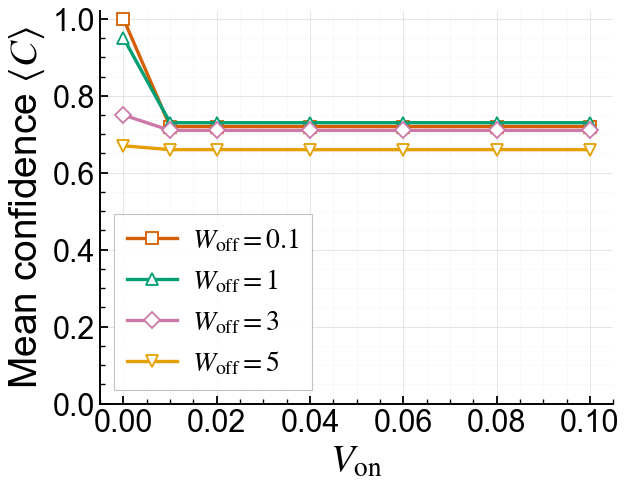}
        \caption{Mean confidence as $\langle C \rangle$ a function of diagonal disorder strength $V_{\text{on}}$}
        \label{mixed1}
    \end{subfigure}
    \hfill
    \begin{subfigure}[t]{0.22\textwidth}
        \centering
        \includegraphics[width=\linewidth]{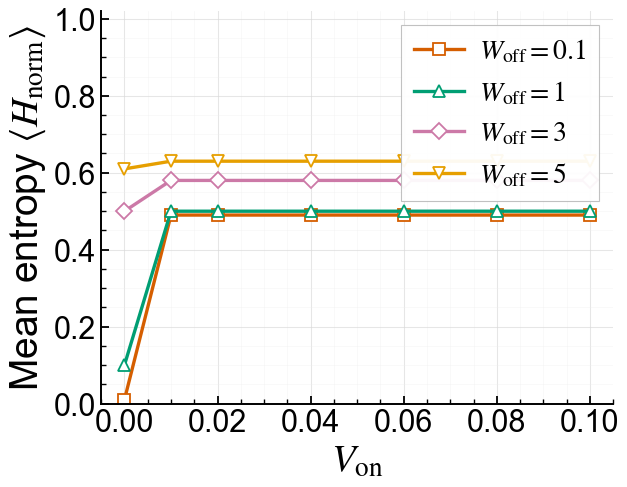}
        \caption{Mean normalized entropy $\langle H_{\text{norm}} \rangle$ as a function of diagonal disorder strength $V_{\text{on}}$}
        \label{mixed2}
    \end{subfigure}
    \caption{Mixed-disorder test of CNN OOD generalization. 
(a) Mean confidence and (b) mean normalized entropy averaged over the \((t_1,t_2)\) phase-diagram region as functions of several fixed off-diagonal disorder strength \(W_{\rm off}\) and varying diagonal disorder strength \(V_{\rm on}\). \(V_{\rm on}=0\) corresponds to chiral-symmetry-preserving disorder, where the CNN remains relatively confident. Increasing \(V_{\rm on}\) suppresses confidence and enhances entropy, indicating that the symmetry-breaking component drives the loss of reliable OOD generalization.}
    \label{fig:mixed_fixed_woff}
\end{figure}

To go beyond the two limiting cases of purely off-diagonal and purely diagonal disorder, we next consider mixed disorder with both \(W_{\rm off}\) and \(V_{\rm on}\) nonzero. This directly tests whether the loss of OOD generalization is controlled by the symmetry-breaking component of the perturbation. We fix \(W_{\rm off}=1.0\) and increase the diagonal disorder strength \(V_{\rm on}\).\\
Figure~\ref{fig:mixed_phase_maps} shows the seed-averaged predicted phase diagrams for \(V_{\rm on}=0.01\), \(0.1\), and \(1.0\). Even for the very weak diagonal component \(V_{\rm on}=0.01\), the clean-system topological phase structure is already strongly destabilized. The predicted map is dominated by fragmented regions and no longer reproduces the well-defined winding-number sectors obtained in the chiral-symmetric case. Increasing \(V_{\rm on}\) further to \(0.1\) and \(1.0\) does not lead to a qualitatively new phase pattern, but mainly enhances the fragmentation and loss of reliable sector boundaries. This mixed-disorder result supports the interpretation that the symmetry-breaking component \(V_{\rm on}\), rather than the total disorder strength alone, controls the loss of CNN OOD generalization. 
This conclusion is further supported by Figure~\ref{fig:mixed_fixed_woff}, which shows the spatially averaged confidence \(\langle C\rangle\) and normalized entropy \(\langle H_{\rm norm}\rangle\) as functions of \(V_{\rm on}\) for several values of \(W_{\rm off}\). For \(V_{\rm on}=0\), the CNN remains highly confident for weak and moderate off-diagonal disorder, consistent with the preservation of chiral symmetry. However, once a small diagonal component is introduced, \(\langle C\rangle\) drops rapidly and \(\langle H_{\rm norm}\rangle\) increases sharply. This change is already pronounced at \(V_{\rm on}=0.01\), after which both quantities vary only weakly with further increasing \(V_{\rm on}\).\\
This behavior indicates that the loss of OOD generalization is mainly triggered by the symmetry-breaking diagonal component rather than by the total disorder strength alone. Strong off-diagonal disorder can also increase the CNN uncertainty by deforming the RCM feature space, as seen from the lower confidence for \(W_{\rm off}=3\) and \(5\) even at \(V_{\rm on}=0\). Nevertheless, the abrupt change upon turning on \(V_{\rm on}\) shows that breaking chiral symmetry qualitatively changes the learned topological representation.\\

\section{PCA Analysis of RCM Feature Space}\label{sec:pca_symmetry}

\begin{figure}[tbp]
    \centering
    \includegraphics[width=0.75\linewidth]{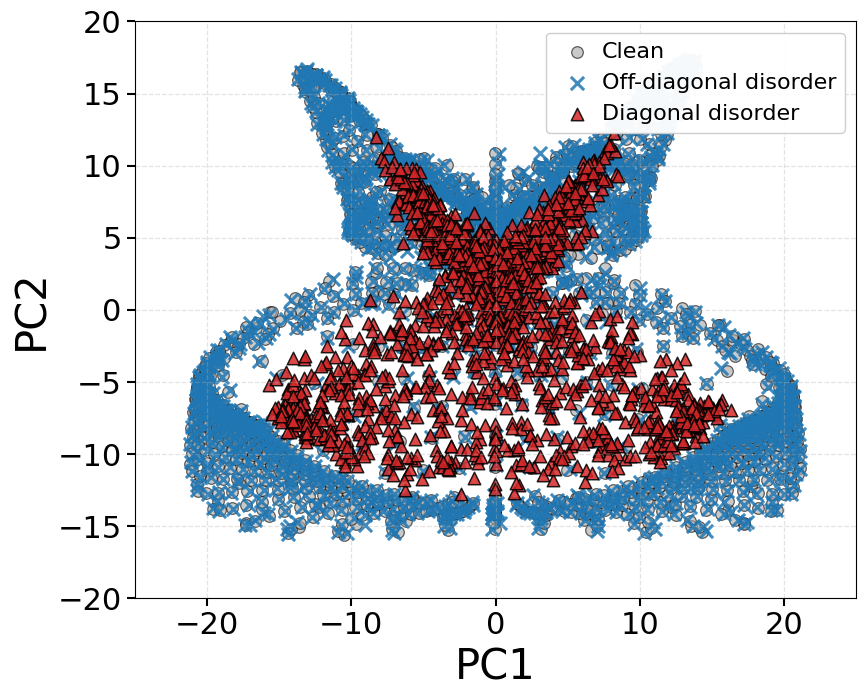}
    \caption{PCA visualizations of RCM spaces. The feature distribution of disorder-free (gray circles) and off-diagonal disorder (blue crosses) samples exhibit substantial overlap in PCA space, while diagonal disorder samples (red circles) diverge.}
    \label{fig:pca}
\end{figure}

To elucidate the physical origin of the CNN's OOD generalization, we analyze the geometry of the RCM feature space under distinct disorder types using PCA\cite{abdi2010principal,wold1987principal,bro2014principal}. RCMs from three ensembles—disorder-free system ($W_{\text{off}} = V_{\text{on}} = 0$), off-diagonal disorder system ($W_{\text{off}} = 0.1$, $V_{\text{on}} = 0$), and diagonal disorder system ($W_{\text{off}} = 0$, $V_{\text{on}} = 0.1$)—are standardized and projected onto the first two principal components, which collectively explain 71.3\% of the total variance (Fig.~\ref{fig:pca}).

The feature distributions of disorder-free and off-diagonal disorder samples exhibit substantial overlap in PCA space, demonstrating that the RCM manifold is topologically invariant under symmetry-preserving perturbations. Despite fluctuations in hopping amplitudes, the topological signatures encoded in the RCM remain robust because the chiral symmetry is preserved. Consequently, the CNN trained on data of disorder-free system can generalize to off-diagonal disordered model.

In contrast, diagonal disorder triggers a transition from topological insulator to Anderson insulator, manifested as a clear geometric separation between disorder-free and disordered samples in feature space, even at $V_{\text{on}} = 0.1$. This geometric shift explains why the CNN's OOD generalization is not applicable in diagonal disordered model.
\section{Edge States, Localization, and Symmetry Breaking}\label{sec6}

In order to shed some light on chiral symmetry preservation manifesting feature overlap while chiral symmetry breaking induces feature space separation, we probe the localization properties of single-particle eigenstates using two complementary diagnostics—IPR and the energy spectrum—across the 
($t_1,t_2$) parameter space. These quantities directly distinguish topological insulator from Anderson insulator.

\subsection{IPR as Spectral Diagnostics}

For a normalized eigenstate $|\phi_{n}\rangle = \sum_i \psi_{n}(i)|i\rangle$ of the Hamiltonian $H$, IPR is defined as:
\begin{equation}
    \text{IPR}(\phi_{n})=\sum_{i=1}^{2N} |\psi_{n}(i)|^4
\end{equation}
where $N=40$ is the number of unit cells. \\
The IPR serves as a robust numerical probe for spatial localization, allowing one to identify localized edge modes and distinguish them from extended bulk states\cite{murphy2011generalized,ghosh2025exploring}. In disordered systems, large IPR values further signal Anderson localization.\cite{calixto2015inverse,evers2000fluctuations,shukla2018disorder,bauer1990correlation,evers2008anderson}.\\
We compute the full energy spectrum $\{E_{n}\}$ and the corresponding IPR values $\{\text{IPR}(\phi_n)\}$ for each disorder realization. All results are averaged over 50 independent realizations to suppress sample-to-sample fluctuations while preserving universal signatures.

\subsection{Energy Spectrum of Disorder-free, Off-diagonal Disorder and Diagonal Disorder Cases}
\begin{figure}[htbp]
    \centering
    \begin{subfigure}[t]{0.15\textwidth}
        \centering
        \includegraphics[width=\linewidth]{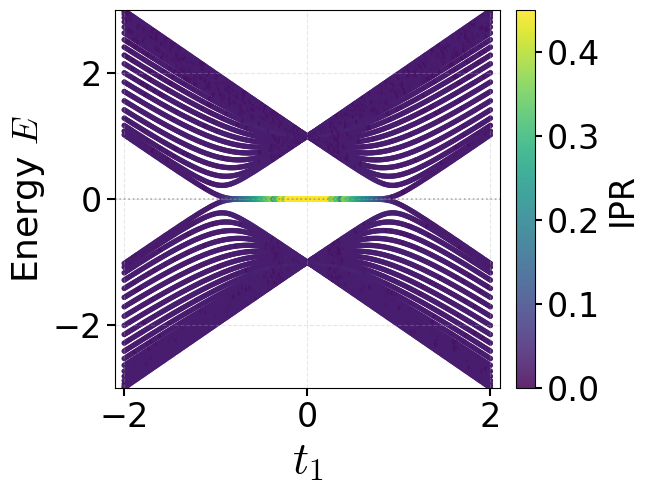}
        \caption{Disorder-free}
        \label{ipr1}
    \end{subfigure}
    \hfill
    \begin{subfigure}[t]{0.15\textwidth}
        \centering
        \includegraphics[width=\linewidth]{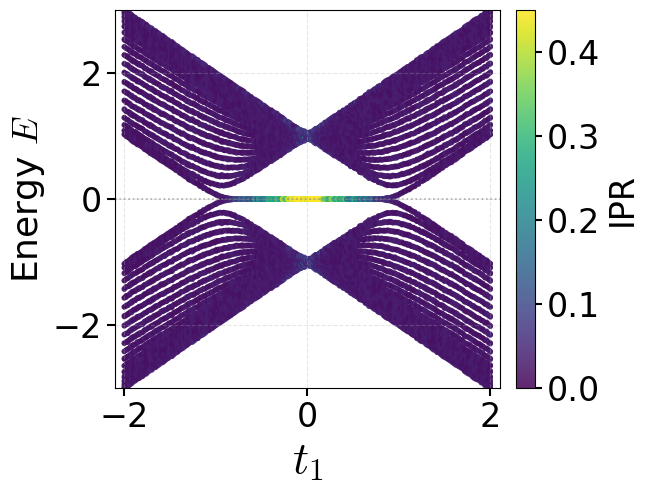}
        \caption{Off-diagonal disorder}
        \label{ipr2}
    \end{subfigure}
    \hfill
    \begin{subfigure}[t]{0.15\textwidth}
        \centering
        \includegraphics[width=\linewidth]{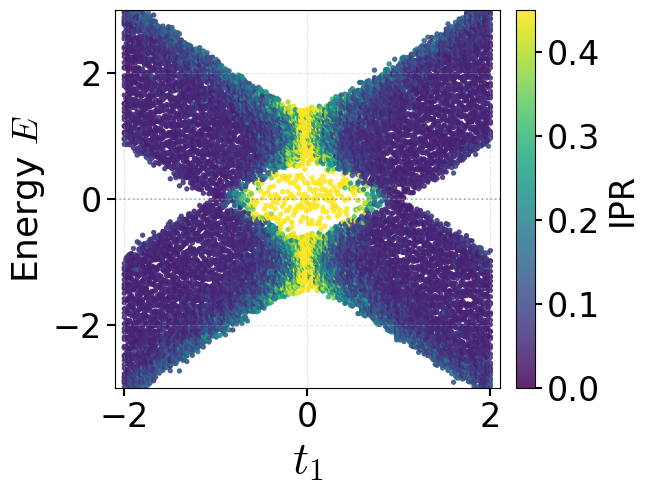}
        \caption{Diagonal disorder}
        \label{ipr3}
    \end{subfigure}
    \caption{Energy spectrum with IPR coloring for SSH model with long-range hopping $t_3 = 1.0$. (a) Disorder-free system, edge states exist, bulk localizations don't exist. (b) Off-diagonal disorder with $W_{\mathrm{off}} = 0.1$, edge states exist, bulk localizations don't exist. (c) Diagonal disorder with $V_{\mathrm{on}} = 0.1$, edge states disappear, bulk localizations appear. The color indicates IPR values.}
    \label{ipr}
\end{figure}

Fig.~\ref{ipr} shows the energy spectrum colored by IPR across the $t_1$ parameter space$(-2<t_1<2)$ at fixed $t_2=0,t_3=1.0$, comparing three cases: disorder-free (a), off-diagonal disorder 
$W_{\text{off}}=0.1$ (b), and diagonal disorder $V_{\text{on}}=0.1$ (c). From Fig.~\ref{exact}, we know that in the disorder-free case, the system exhibits a topological phase with winding number $\nu \neq 0$ for $|t_1|<1$, while for $|t_1|>1$ , the system transitions to a trivial state ($\nu=0$).

In the disorder-free case (Fig.~\ref{ipr1}), degenerate zero-energy modes ($E=0$) emerge precisely within the topological region ($-1.0 < t_1 < 1.0$). These modes exhibit IPR values significantly higher than the bulk states, characteristic of well-defined edge states. Under off-diagonal disorder (Fig.~\ref{ipr2}), these zero-energy modes persist robustly across the topological phase. Their energy remains pinned at the Fermi level ($E=0$) and their IPR remains high, confirming that the system remains a topological insulator with its chiral-symmetry-protected topology surviving despite disorder-induced bulk spectral fluctuations \cite{murphy2011generalized, evers2000fluctuations}.\\
In contrast, diagonal disorder (Fig.~\ref{ipr3}) destroys the topological signatures even at weak strengths. The previously degenerate edge modes are shifted from zero energy and hybridize with the bulk. Notably, the IPR increases compared with disorder-free system and off-diagonal disordered system, signaling a transition where the system's localization mechanism shifts from topological edge protection to Anderson localization across the entire spectrum. This spectral collapse and the loss of mid-gap states provide the structural origin for the PCA manifold separation discussed in Sec.~\ref{sec:pca_symmetry}: once chiral symmetry is broken, the RCM loses the distinct boundary-bulk correlations that define the topological manifold. Consequently, the CNN's OOD generalizations is not applicable in diagonal disordered model with chiral symmetry broken because this system has undergone a phase transition into an Anderson insulator.

\subsection{Energy Spectrum of Different Off-diagonal and Diagonal Disorder Strengths}
\begin{figure}[htbp]
    \centering
    \begin{subfigure}{0.22\textwidth}
        \centering
        \includegraphics[width=\linewidth]{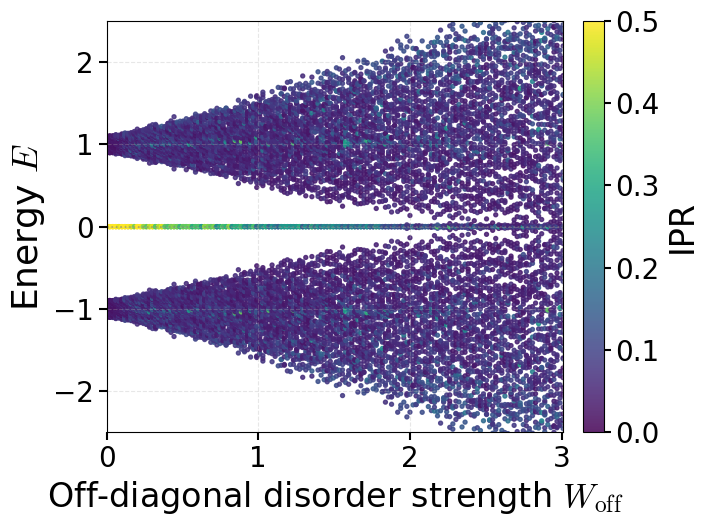}
        \caption{Off-diagonal disorder}
        \label{ipr4}
    \end{subfigure}
    \hfill
    \begin{subfigure}{0.22\textwidth}
        \centering
        \includegraphics[width=\linewidth]{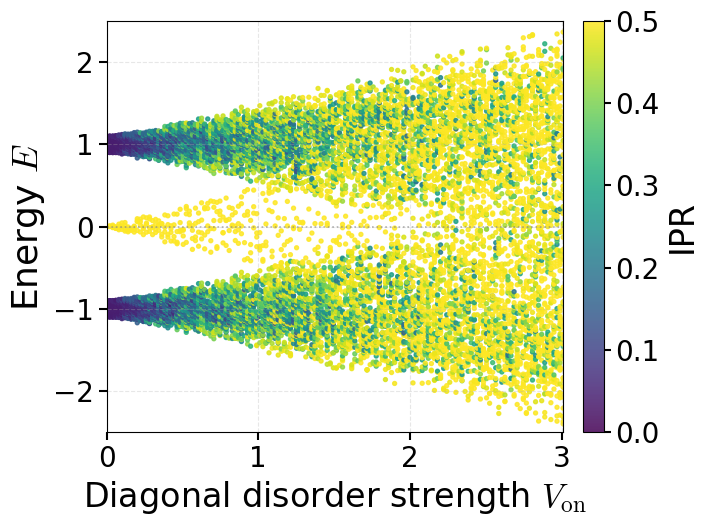}
        \caption{Diagonal disorder}
        \label{ipr5}
    \end{subfigure}
    \caption{Effect of disorder on energy spectrum and IPR ($t_1 = 0.1$, $t_2 = 0.0$, $t_3 = 1.0$). (a) Off-diagonal disorder $(W_{\mathrm{off}})$. Edge states exist persistently, bulk states are not localized. (b) Diagonal disorder $(V_{\mathrm{on}})$. Edge states disappear, bulk states become localized. The color indicates IPR values.}
    \label{fig:4_2}
\end{figure}

To quantify the robustness of edge states, Fig.\ref{fig:4_2} tracks spectral evolution at a fixed topological point ($t_1=0.1,t_2=0,t_3=1.0$) as disorder strength increases.\\ Under off-diagonal disorder (Fig.~\ref{ipr4}), the zero-energy modes remain pinned at $E=0$ even if the disorder strength becomes larger, with bulk states' IPR staying low throughout. This confirms that the system remains a topological insulator with its edge-state signatures intact.\\
Under diagonal disorder (Fig.~\ref{ipr5}), however, even infinitesimal $V_{\text{on}}$ triggers immediate shift of the zero-energy modes. As $V_{\text{on}}$ increases, the entire spectrum undergoes a disorder-driven localization transition. By $V_{\text{on}} \approx 2.0$, all states exhibit high IPR values, signaling that the system has fully evolved from a topological phase into an Anderson insulating phase \cite{li2009topological, groth2009theory,xing2011topological}.This topological collapse at arbitrarily weak symmetry breaking disorder provides the sturctural explanation for the CNN's catastrophic failure discussed in Sec.~\ref{sec4}. Because the RCM's structure is fundamentally tethered to these symmetry-protected edge states, their immediate dissolution under diagonal disorder means the CNN is presented with data that no longer contains the "topological fingerprints" it was trained to recognize. Therefore, the CNN's failure is not a limitation of the model architecture, but a direct reflection of the fundamental breakdown of symmetry-protected features at the microscopic level.

\section{Conclusion}

In summary, our work establishes that the OOD generalization of machine learning in topological systems is fundamentally governed by the underlying physical symmetry. We show that the machine learning achieves seamless OOD success as long as chiral symmetry is preserved, maintaining the system as a topological insulator; conversely, it fails when chiral symmetry is broken, as the system transitions into an Anderson insulator.

PCA shows that symmetry preservation ensures an overlap of feature manifolds, while symmetry breaking induces the separation. Finally, IPR and energy spectrum analysis provides the structural evidence, confirming that the "learning failure" is a direct reflection of the dissolution of topological edge states into localized bulk states. These results suggest that clean-trained CNNs can provide useful data-driven diagnostics for disordered topological systems, provided that their predictions are interpreted together with feature-space and localization analyses. In the extended SSH chain studied here, the success or failure of OOD generalization reflects whether the RCM representation remains connected to the clean topological feature manifold or is disrupted by chiral-symmetry breaking and Anderson localization.

\appendix
\section{Robustness with respect to CNN initialization}\label{app:seed_robustness}
To test the robustness of the CNN predictions against random initialization, we train \(N_{\rm seed}=10\) independent CNNs and evaluate \(K=50\) disorder realizations for each parameter point. For each case, the softmax probabilities are first averaged over disorder realizations and network seeds to obtain the phase diagrams. The confidence \(C=\max_\nu \bar p_\nu\) and normalized entropy \(H_{\rm norm}=-\sum_\nu \bar p_\nu \log \bar p_\nu/\log 3\) are then averaged over the \((t_1,t_2)\) region shown in the phase diagrams to get the mean values $\langle C \rangle$ and $\langle H_{\text{norm}}\rangle$. Table~\ref{tab:seed_robustness} also reports the mean seed-to-seed standard deviations $\langle\sigma_\text{seed}(C)\rangle$ and $\langle\sigma_\text{seed}(H_{\text{norm}})\rangle$, computed at each parameter point and then averaged over the same region.\\
As shown in Table~\ref{tab:seed_robustness}, the CNN predictions are statistically robust over independent trainings. For weak off-diagonal disorder, the mean confidence is close to unity and the normalized entropy is nearly zero. As \(W_{\rm off}\) increases from \(0.1\) to \(5.0\), the confidence gradually decreases from \(0.9967\) to \(0.7233\), while the entropy increases from \(0.0091\) to \(0.5613\). The corresponding seed-to-seed fluctuations also increase with disorder strength, but remain moderate. This confirms that the observed OOD behavior under off-diagonal disorder is not an artifact of a particular CNN initialization.\\
For diagonal disorder, even a small value \(V_{\rm on}=0.1\) already gives \(\langle C\rangle=0.7298\) and \(\langle H_{\rm norm}\rangle=0.5466\), comparable to the strong off-diagonal disorder case \(W_{\rm off}=5.0\). The seed-to-seed fluctuations are also of similar magnitude. This comparison shows that a weak symmetry-breaking diagonal perturbation can destabilize the clean-trained CNN to an extent comparable to much stronger symmetry-preserving off-diagonal disorder, supporting the conclusion that the symmetry class of the perturbation plays a key role in OOD generalization.
\begin{table}[h]
\caption{
Seed-to-seed robustness of CNN predictions. 
Here \(\langle C\rangle\), \(\langle H_{\rm norm}\rangle\), 
\(\langle \sigma_{\rm seed}(C)\rangle\), and 
\(\langle \sigma_{\rm seed}(H_{\rm norm})\rangle\) denote the spatially averaged confidence, normalized entropy, and their mean seed-to-seed standard deviations, respectively. 
For off-diagonal disorder, the confidence decreases and the entropy increases gradually as \(W_{\rm off}\) increases, while the seed-to-seed fluctuations remain moderate, showing that the CNN predictions are reproducible across independent trainings. 
For diagonal disorder, even a weak symmetry-breaking perturbation \(V_{\rm on}=0.1\) produces confidence, entropy, and seed-to-seed fluctuations comparable to those of the much stronger off-diagonal disorder case \(W_{\rm off}=5.0\), indicating that diagonal disorder destabilizes the clean-trained CNN more efficiently than symmetry-preserving off-diagonal disorder.
}
\label{tab:seed_robustness}
\begin{ruledtabular}
\begin{tabular}{lcccc}
Case 
& \(\langle C\rangle\) 
& \(\langle H_{\rm norm}\rangle\) 
& \(\langle \sigma_{\rm seed}(C)\rangle\) 
& \(\langle \sigma_{\rm seed}(H_{\rm norm})\rangle\) \\
\hline
\(W_{\rm off}=0.1\) & 0.9967 & 0.0091 & 0.0036 & 0.0071 \\
\(W_{\rm off}=1.0\) & 0.9646 & 0.0687 & 0.0127 & 0.0215 \\
\(W_{\rm off}=2.0\) & 0.9010 & 0.1984 & 0.0480 & 0.0807 \\
\(W_{\rm off}=3.0\) & 0.7998 & 0.3870 & 0.0903 & 0.1450 \\
\(W_{\rm off}=4.0\) & 0.7534 & 0.5023 & 0.1223 & 0.1986 \\
\(W_{\rm off}=5.0\) & 0.7233 & 0.5613 & 0.1394 & 0.2256 \\
\(V_{\rm on}=0.1\) & 0.7298 & 0.5466 & 0.1385 & 0.2383 \\
\end{tabular}
\end{ruledtabular}
\end{table}

\section{Supplementary Plots for Other $W_{\text{off}}$ Values}\label{App2}

\begin{figure}[h]
    \centering
    \tiny
    
    \begin{subfigure}[t]{0.15\textwidth}
        \centering
        \includegraphics[width=0.9\textwidth]{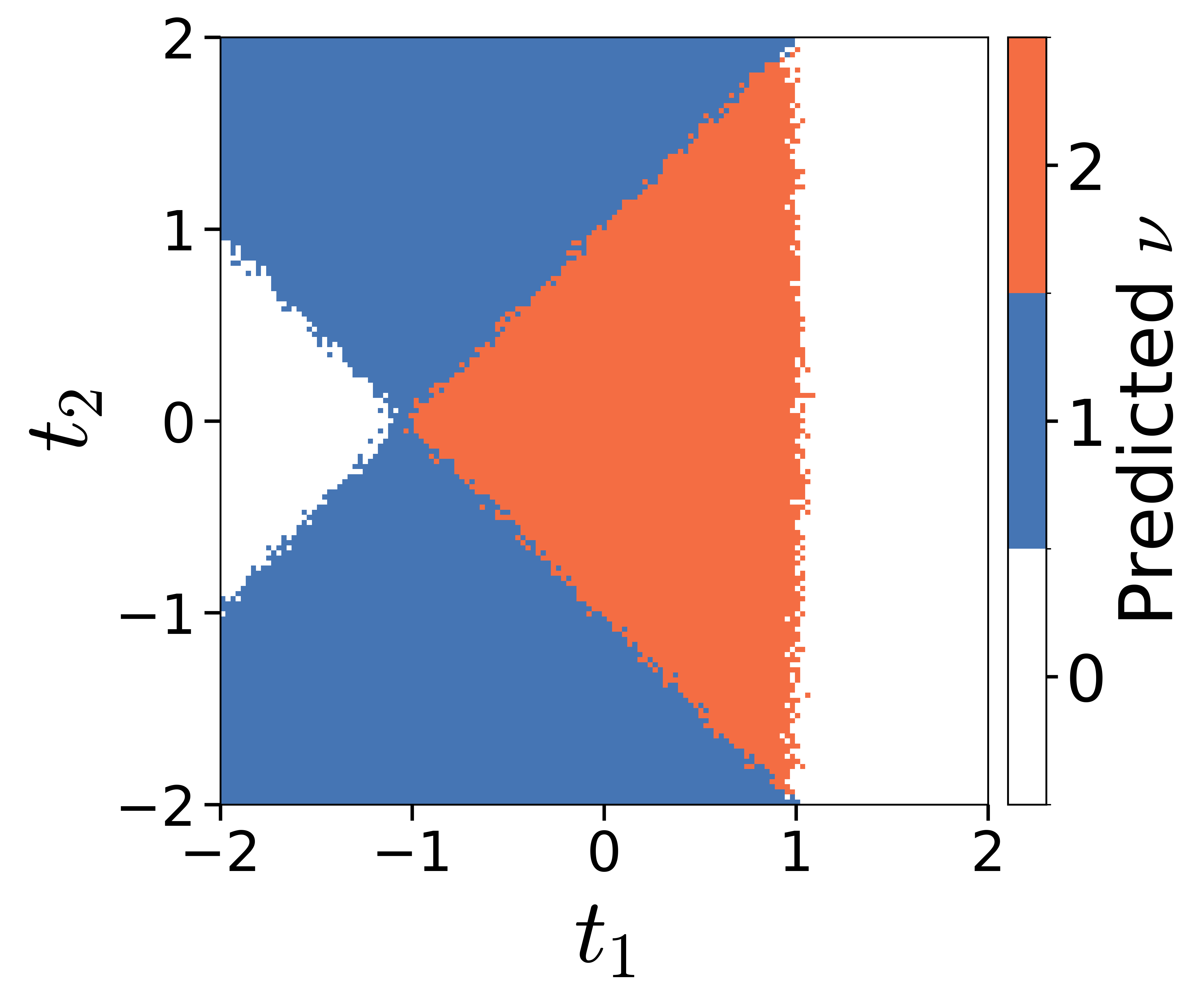}
        \vspace{-3mm}
        \caption*{\footnotesize$\nu$ ($W_{\text{off}}=1.0$)}
        \label{fig:2a}
    \end{subfigure}
    \hspace{0.2mm}
    \begin{subfigure}[t]{0.15\textwidth}
        \centering
        \includegraphics[width=0.9\textwidth]{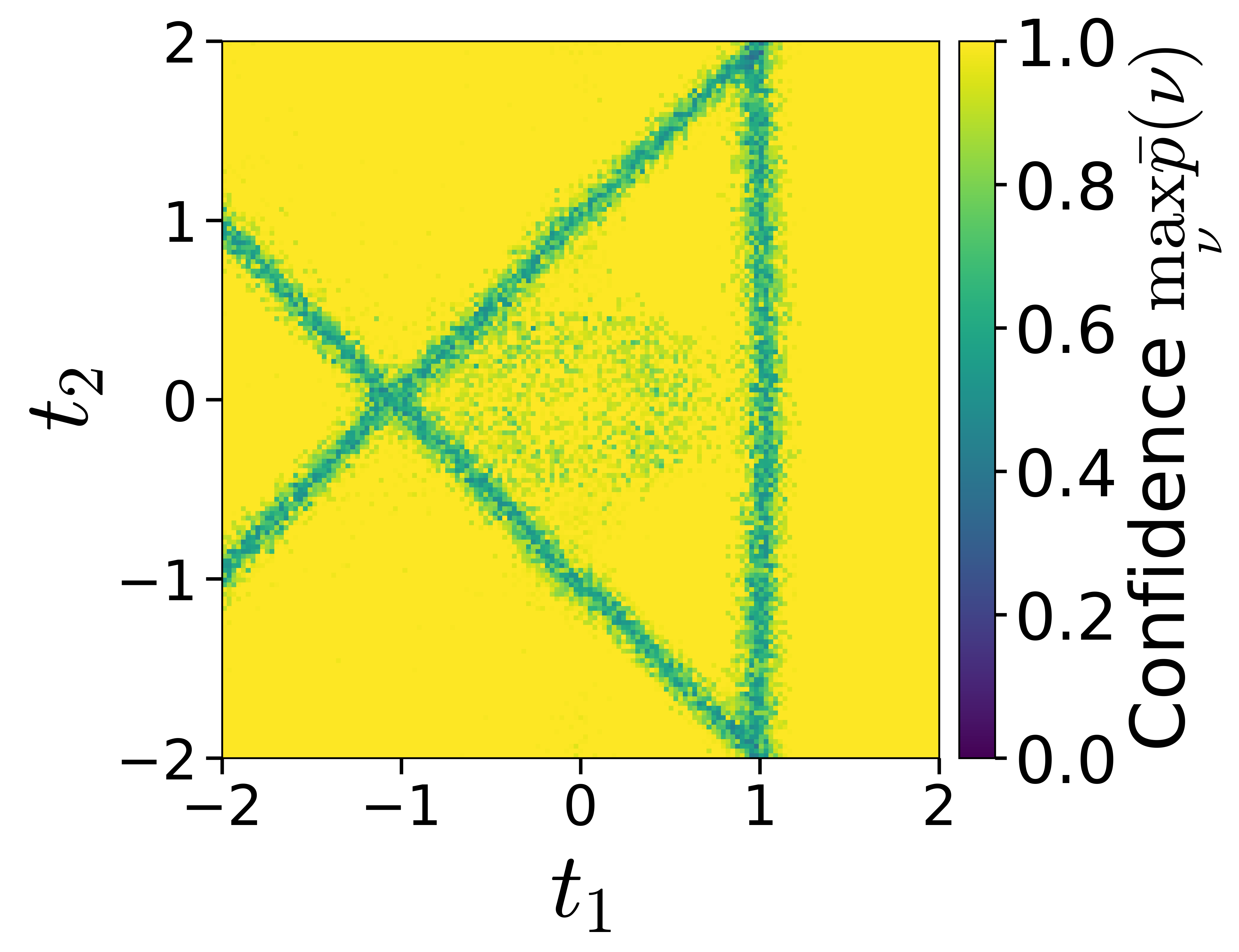}
        \vspace{-3mm}
        \caption*{\footnotesize Confidence}
        \label{fig:2b}
    \end{subfigure}
    \hspace{0.2mm}
    \begin{subfigure}[t]{0.15\textwidth}
        \centering
        \includegraphics[width=0.9\textwidth]{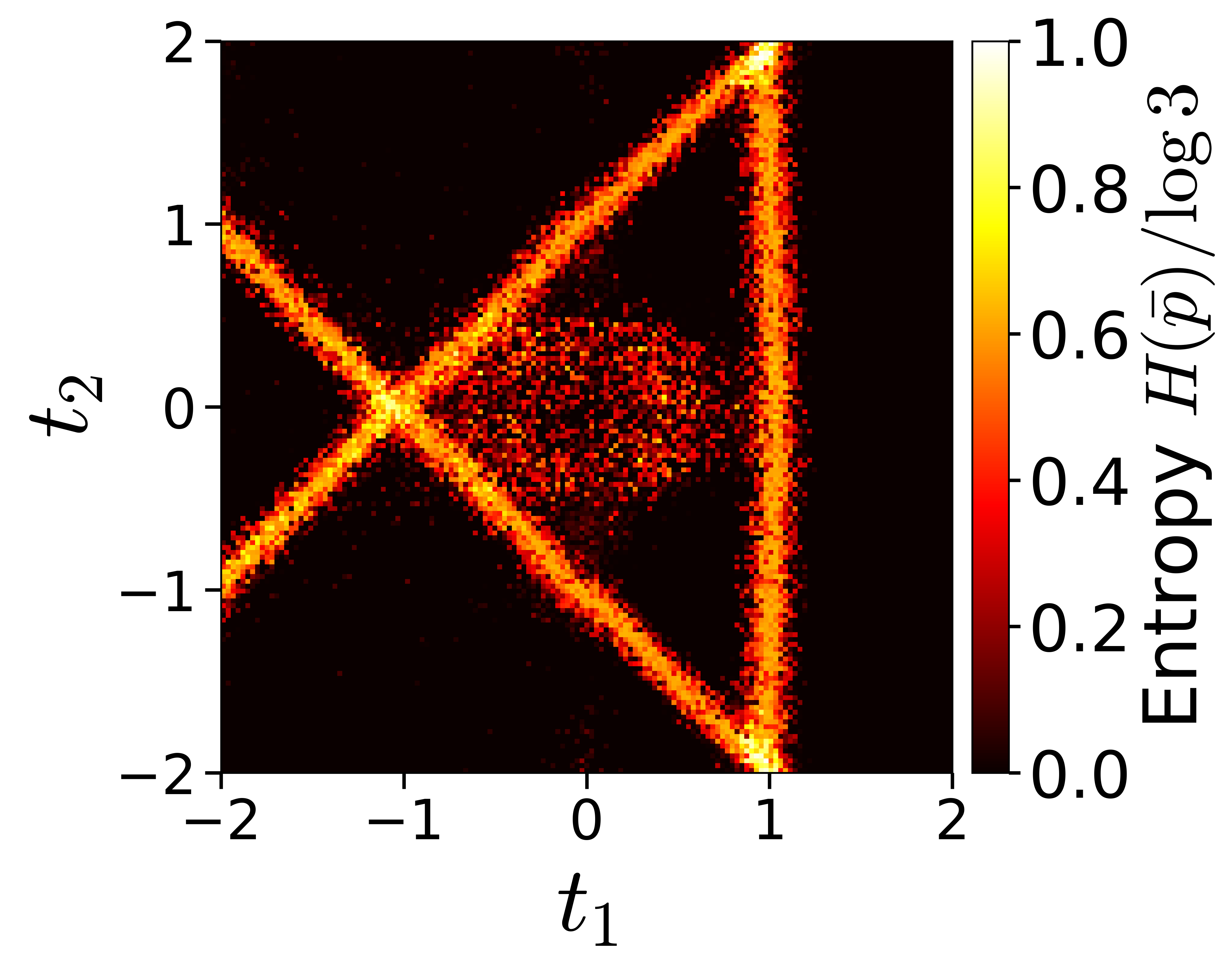}
        \vspace{-3mm}
        \caption*{\footnotesize Entropy}
        \label{fig:2c}
    \end{subfigure}
    
    \vspace{0.3mm}
    
    \begin{subfigure}[t]{0.15\textwidth}
        \centering
        \includegraphics[width=0.9\textwidth]{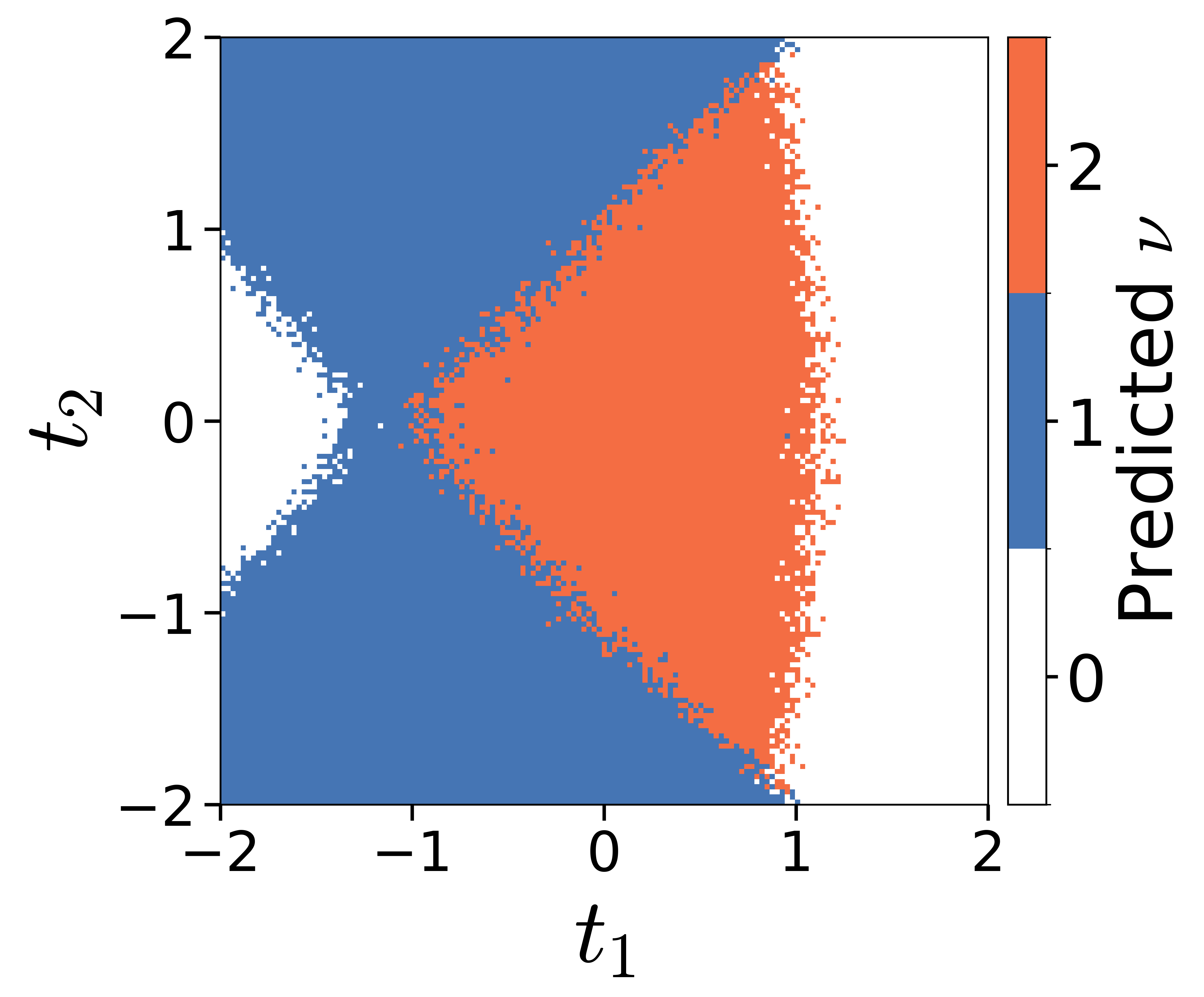}
        \vspace{-3mm}
        \caption*{\footnotesize$\nu$ ($W_{\text{off}}=2.0$)}
        \label{fig:3a}
    \end{subfigure}
    \hspace{0.2mm}
    \begin{subfigure}[t]{0.15\textwidth}
        \centering
        \includegraphics[width=0.9\textwidth]{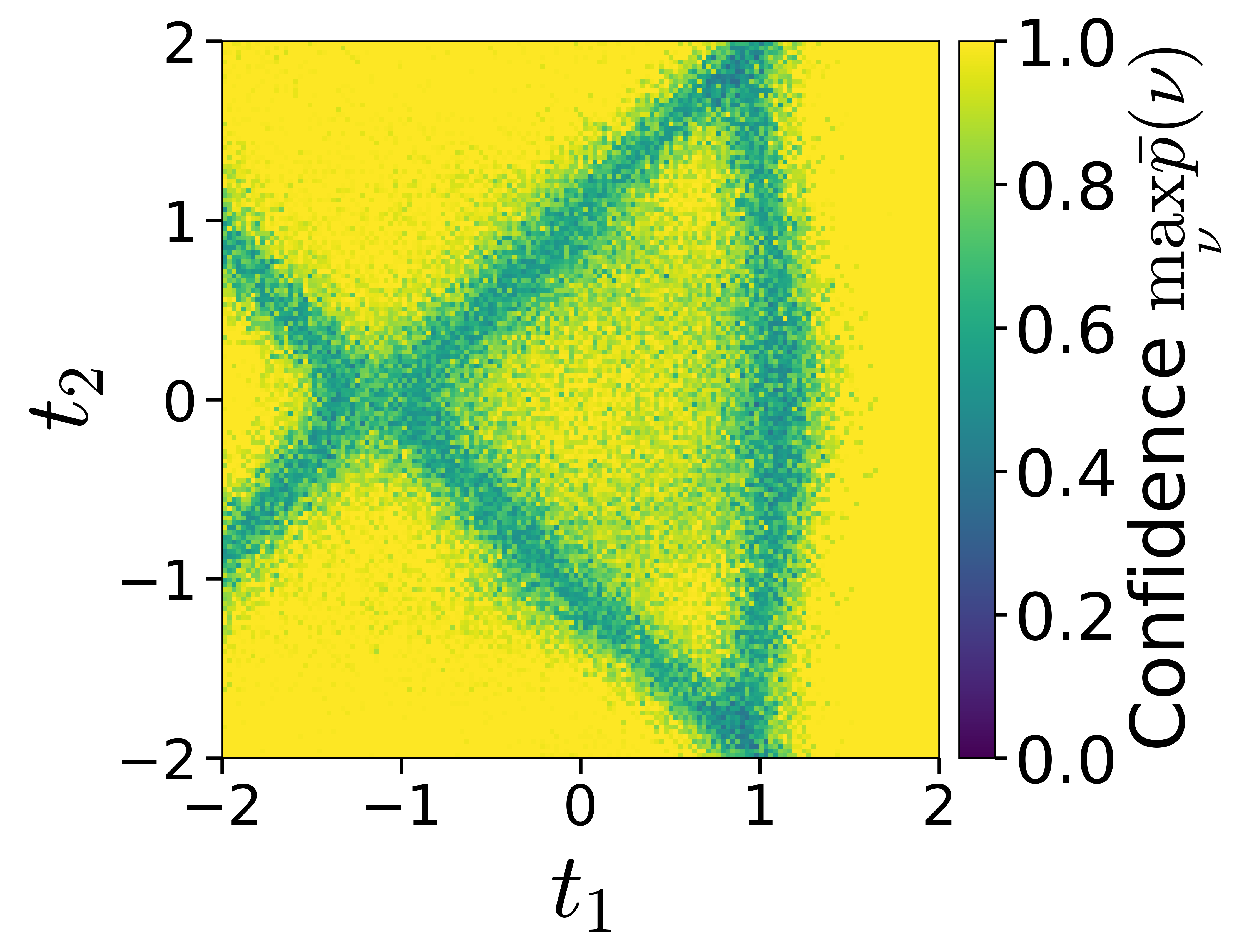}
        \vspace{-3mm}
        \caption*{\footnotesize Confidence}
        \label{fig:3b}
    \end{subfigure}
    \hspace{0.2mm}
    \begin{subfigure}[t]{0.15\textwidth}
        \centering
        \includegraphics[width=0.9\textwidth]{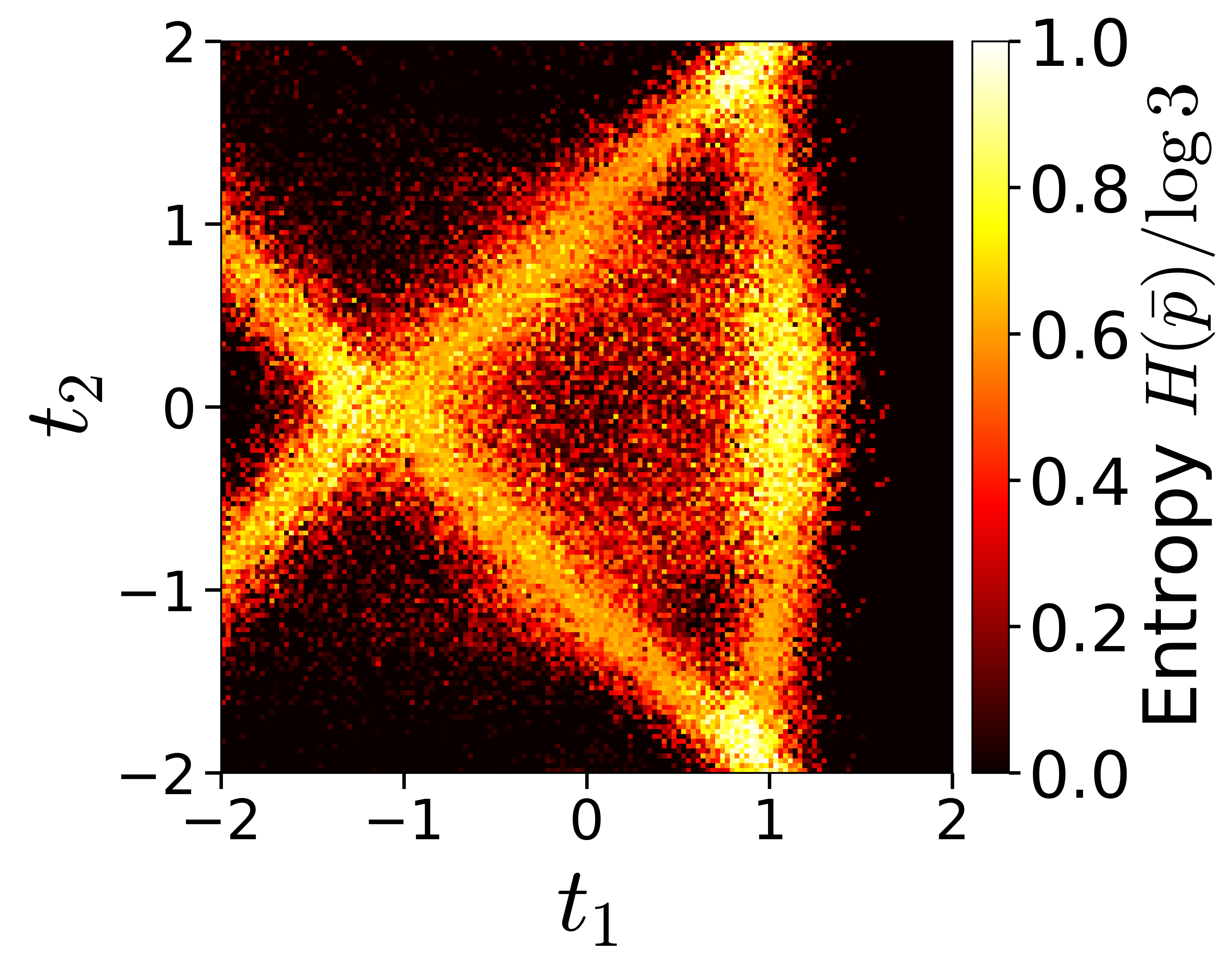}
        \vspace{-3mm}
        \caption*{\footnotesize Entropy}
        \label{fig:3c}
    \end{subfigure}
    
    \vspace{0.3mm}
    
    \begin{subfigure}[t]{0.15\textwidth}
        \centering
        \includegraphics[width=0.9\textwidth]{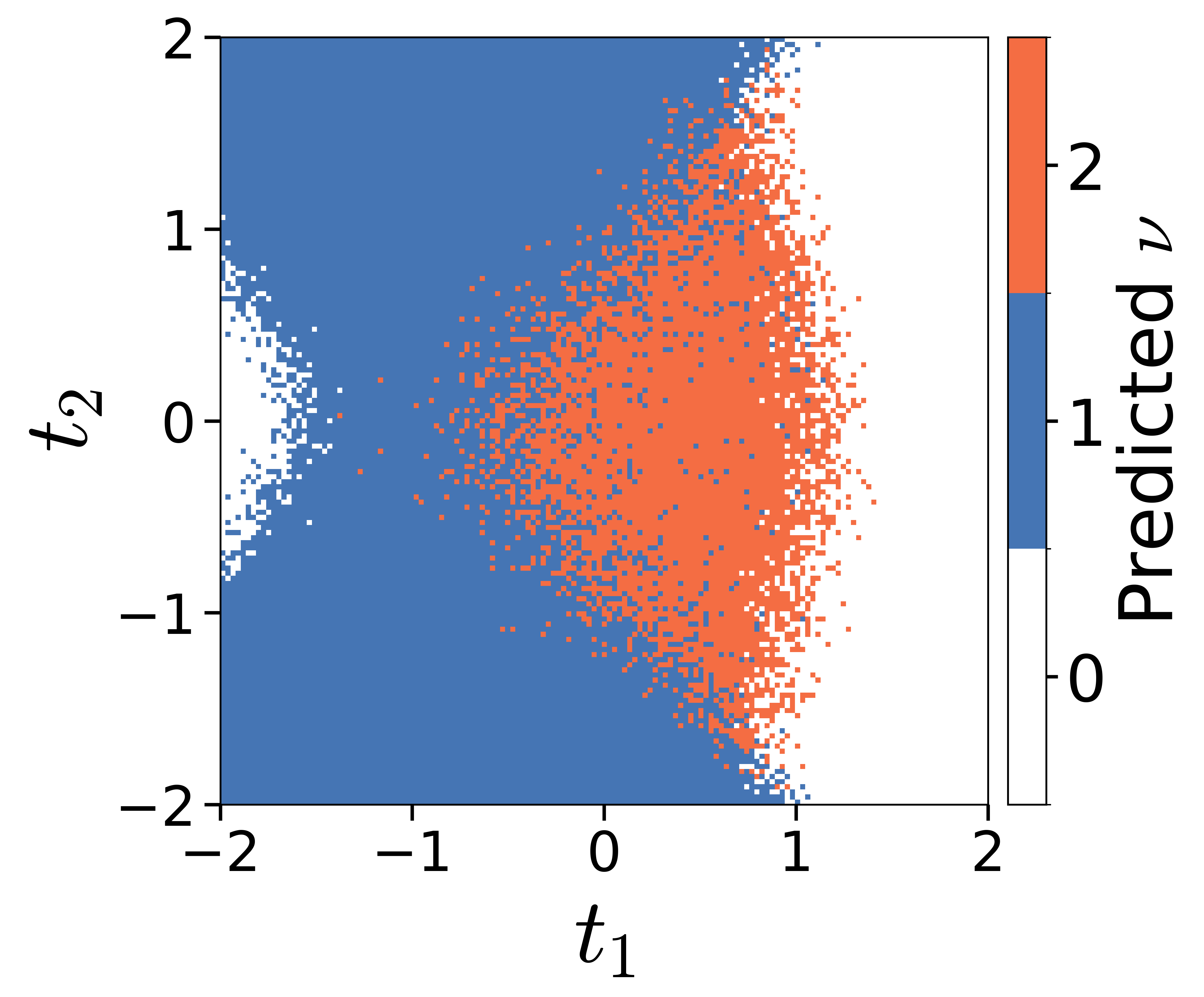}
        \vspace{-3mm}
        \caption*{\footnotesize$\nu$ ($W_{\text{off}}=3.0$)}
        \label{fig:4a}
    \end{subfigure}
    \hspace{0.2mm}
    \begin{subfigure}[t]{0.15\textwidth}
        \centering
        \includegraphics[width=0.9\textwidth]{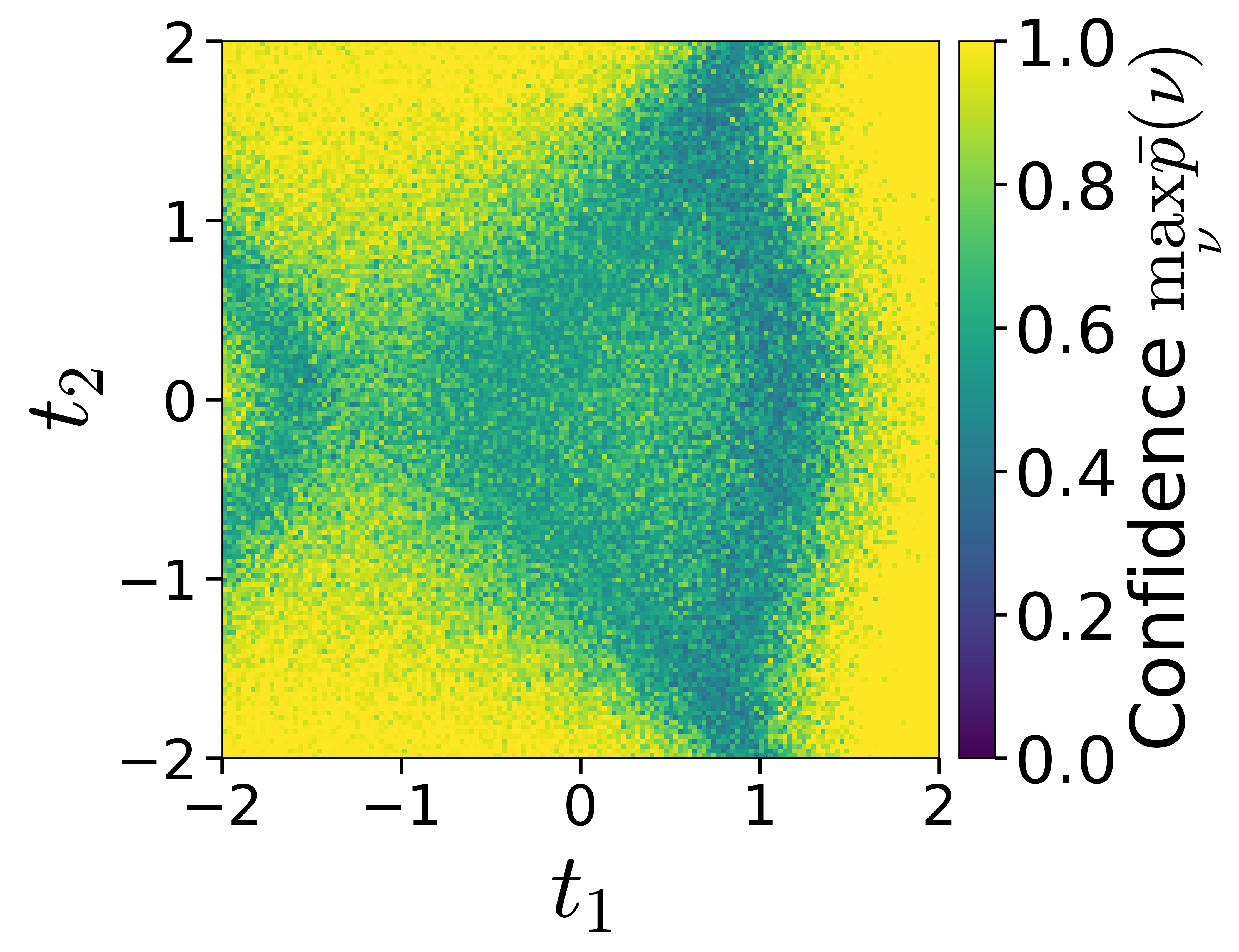}
        \vspace{-3mm}
        \caption*{\footnotesize Confidence}
        \label{fig:4b}
    \end{subfigure}
    \hspace{0.2mm}
    \begin{subfigure}[t]{0.15\textwidth}
        \centering
        \includegraphics[width=0.9\textwidth]{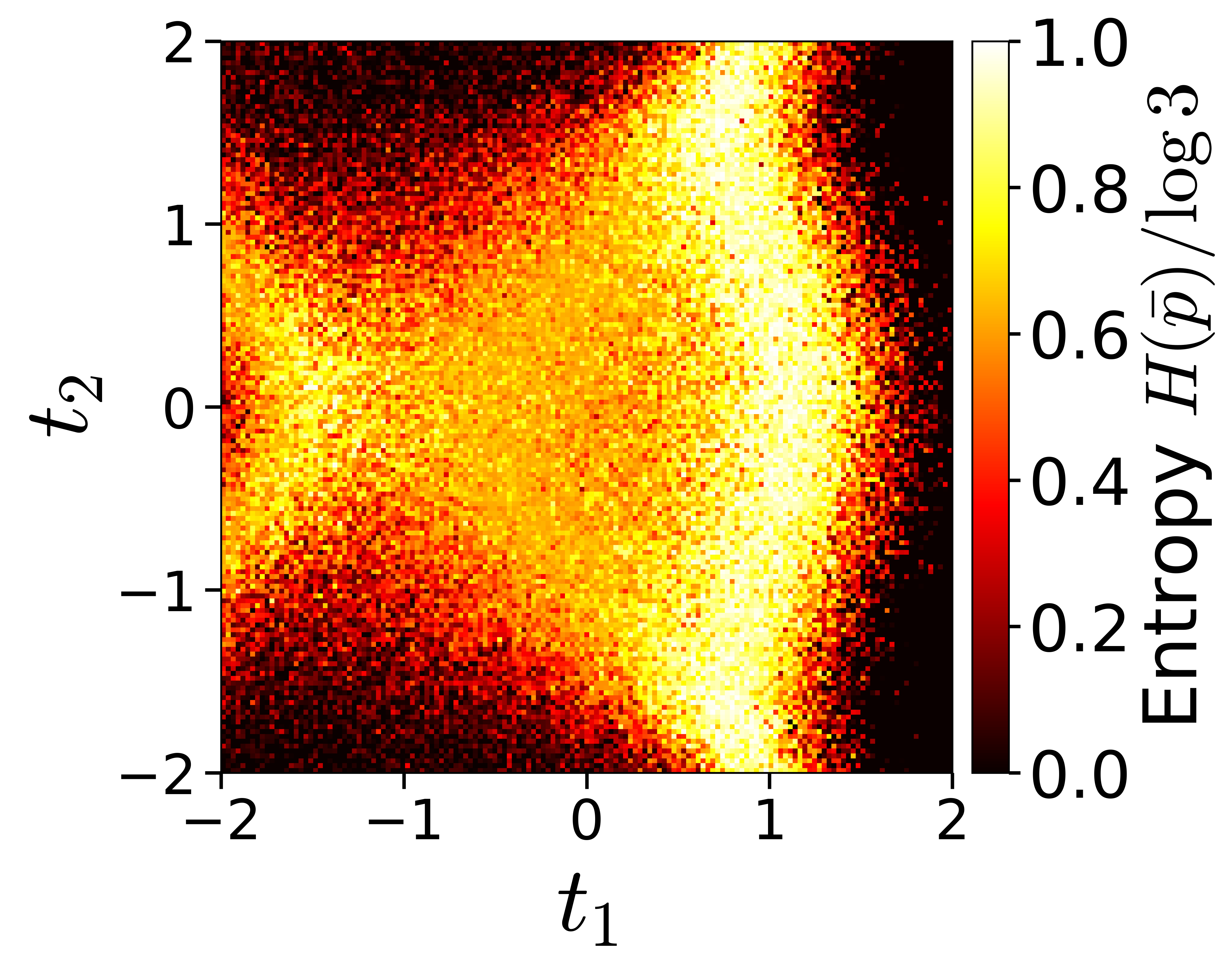}
        \vspace{-3mm}
        \caption*{\footnotesize Entropy}
        \label{fig:4c}
    \end{subfigure}
    
    \vspace{0.3mm}
    
    \begin{subfigure}[t]{0.15\textwidth}
        \centering
        \includegraphics[width=0.9\textwidth]{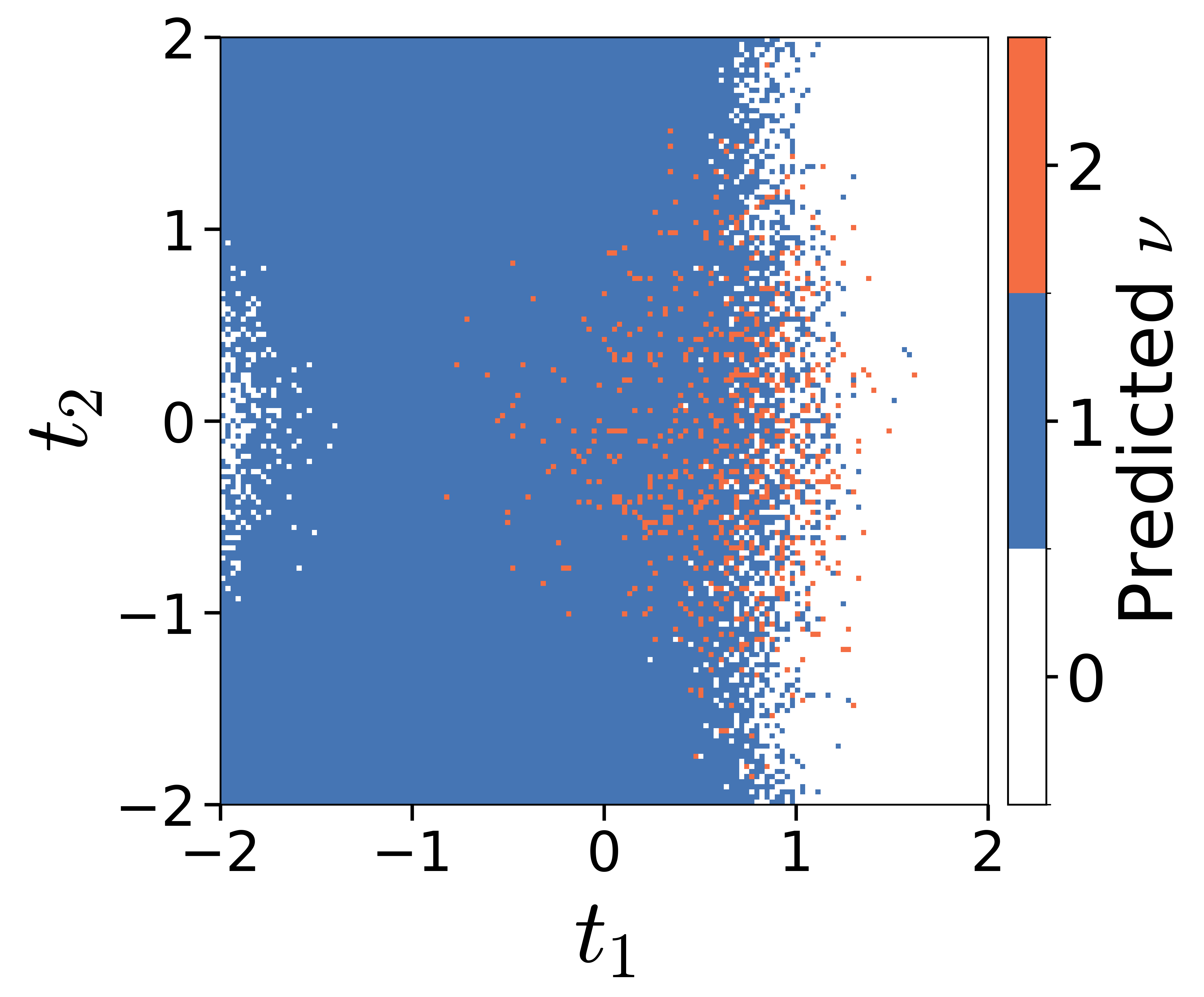}
        \vspace{-3mm}
        \caption*{\footnotesize$\nu$ $W_{\text{off}}=4.0$}
        \label{fig:5a}
    \end{subfigure}
    \hspace{0.2mm}
    \begin{subfigure}[t]{0.15\textwidth}
        \centering
        \includegraphics[width=0.9\textwidth]{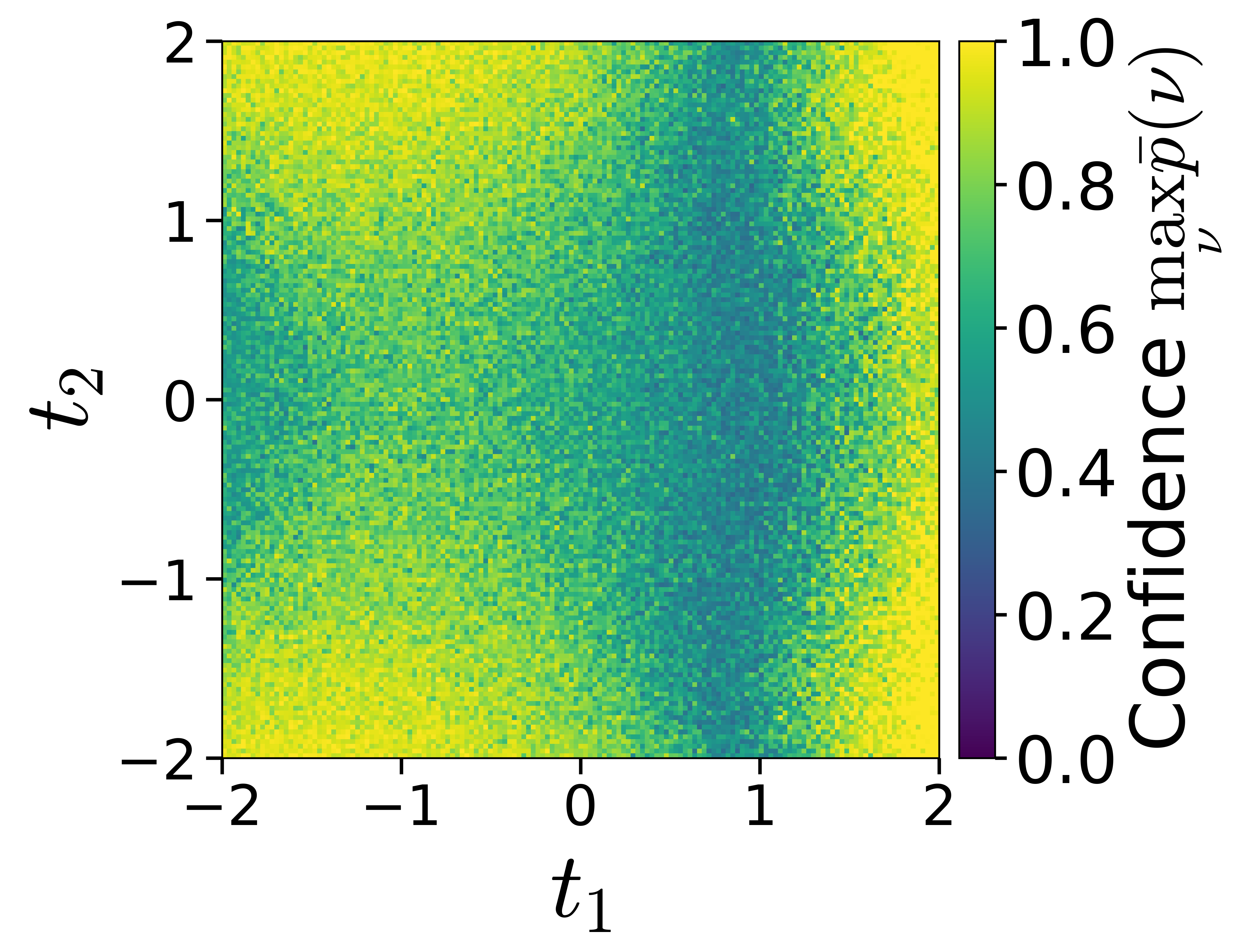}
        \vspace{-3mm}
        \caption*{\footnotesize Confidence}
        \label{fig:5b}
    \end{subfigure}
    \hspace{0.2mm}
    \begin{subfigure}[t]{0.15\textwidth}
        \centering
        \includegraphics[width=0.9\textwidth]{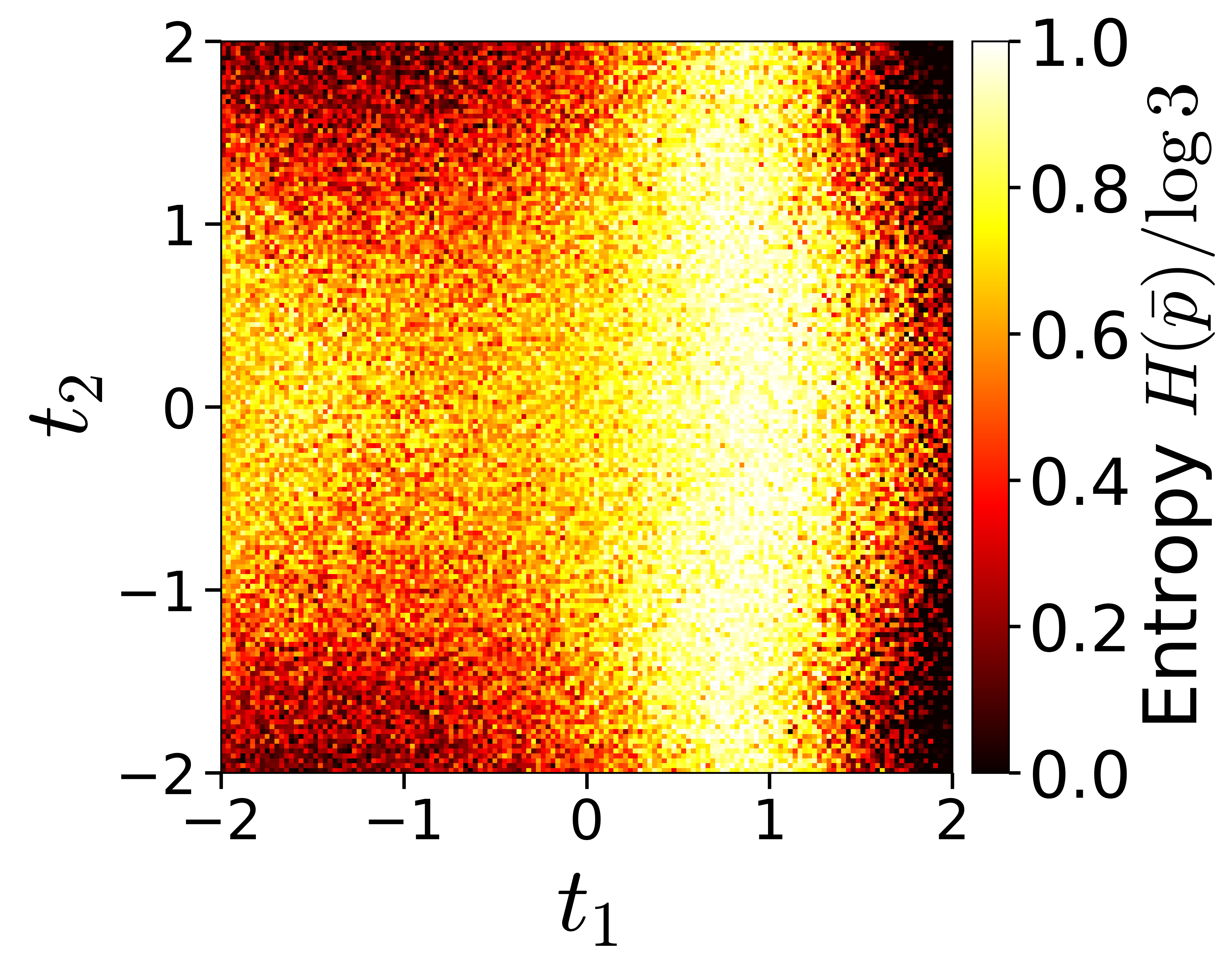}
        \vspace{-3mm}
        \caption*{\footnotesize Entropy}
        \label{fig:5c}
    \end{subfigure}
    
    \vspace{0.3mm}
    
    \begin{subfigure}[t]{0.15\textwidth}
        \centering
        \includegraphics[width=0.9\textwidth]{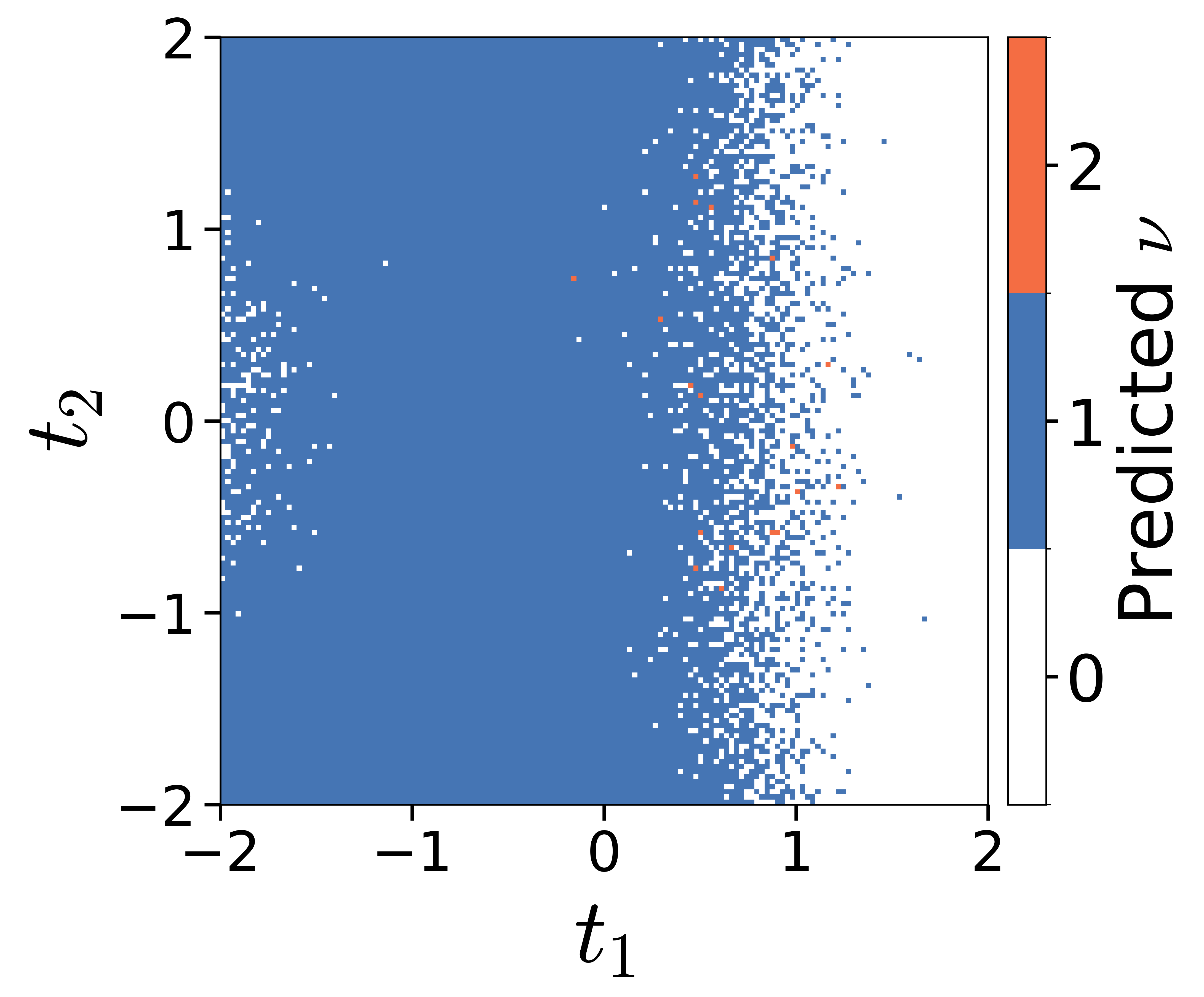}
        \vspace{-3mm}
        \caption*{\footnotesize$\nu$ ($W_{\text{off}}=5.0$)}
        \label{fig:6a}
    \end{subfigure}
    \hspace{0.2mm}
    \begin{subfigure}[t]{0.15\textwidth}
        \centering
        \includegraphics[width=0.9\textwidth]{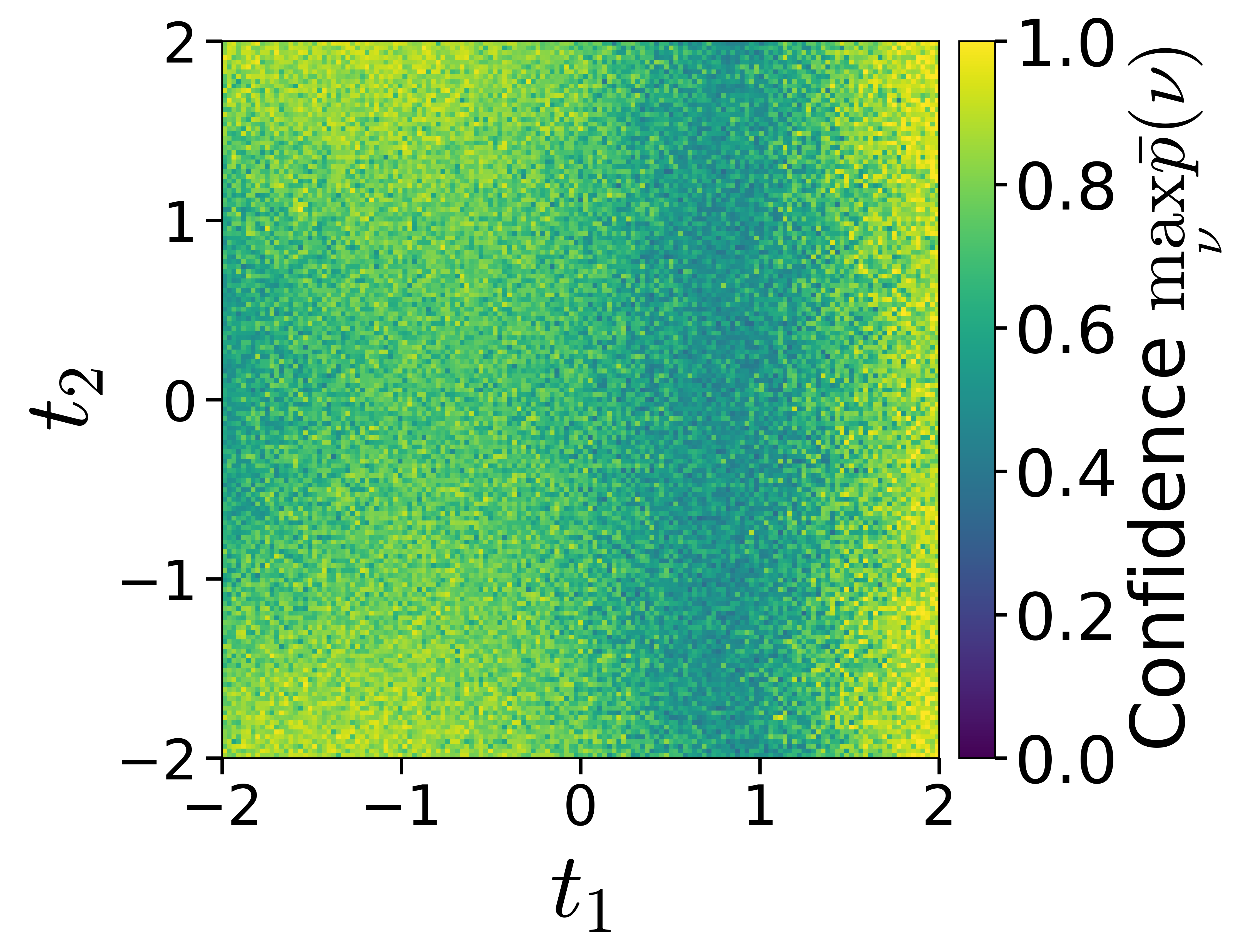}
        \vspace{-3mm}
        \caption*{\footnotesize Confidence}
        \label{fig:6b}
    \end{subfigure}
    \hspace{0.2mm}
    \begin{subfigure}[t]{0.15\textwidth}
        \centering
        \includegraphics[width=0.9\textwidth]{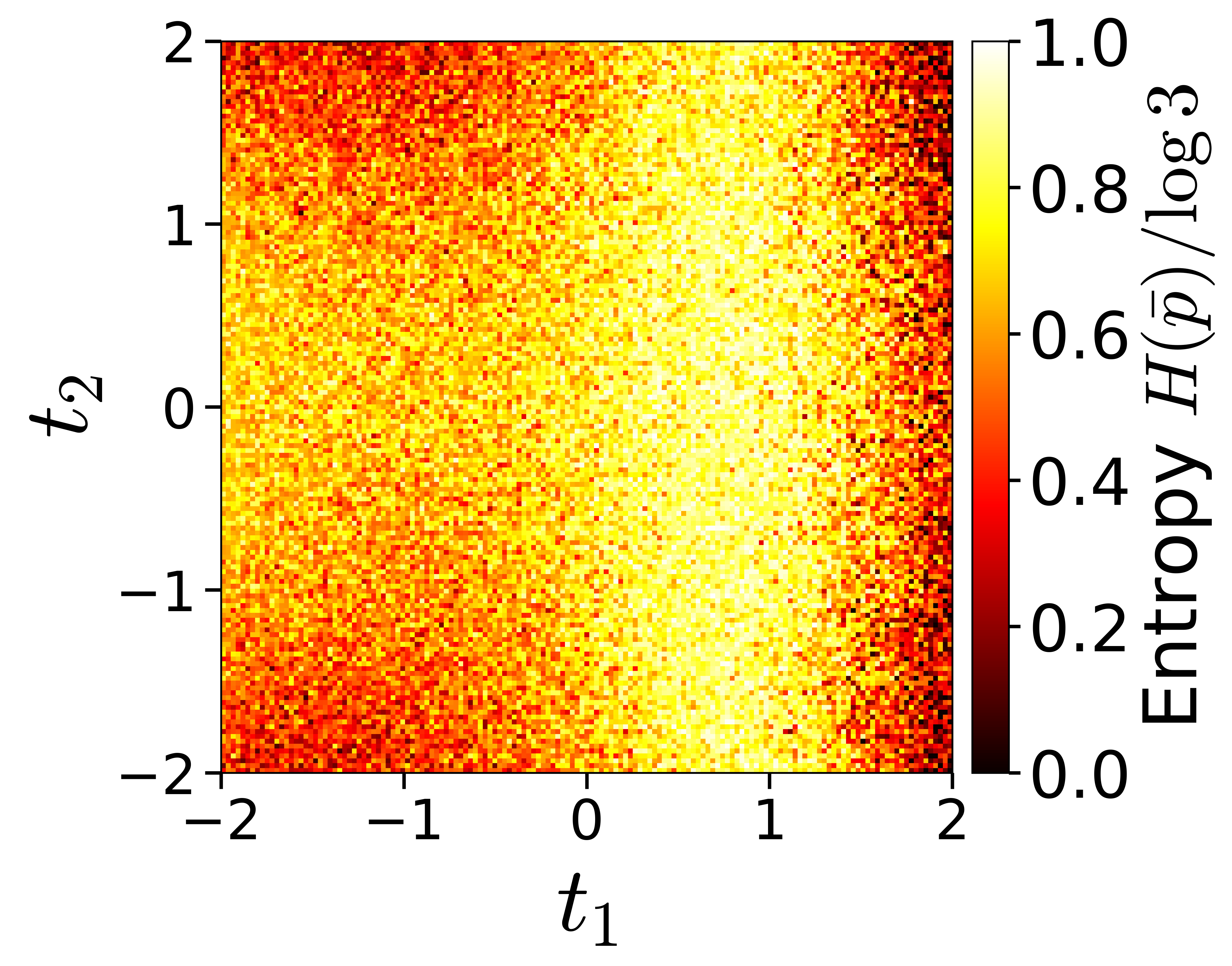}
        \vspace{-3mm}
        \caption*{\footnotesize Entropy}
        \label{fig:6c}
    \end{subfigure}
    \caption{Prediction results for varying disorder strengths $(W_{\text{off}})$. From top to bottom: 
row 1: $W_{\text{off}}=1.0$, row 2: $W_{\text{off}}=2.0$, 
row 3: $W_{\text{off}}=3.0$, row 4: $W_{\text{off}}=4.0$, 
row 5: $W_{\text{off}}=5.0$. 
From left to right: (1) predicted winding number maps, as $W_{\text{off}}$ increases, boundaries gradually blur, but the topological structure remains accurate until $W_{\text{off}}\approx3.0$, 
(2) confidence maps indicating prediction certainty, as $W_{\text{off}}$ increase, confidence decreases, 
(3) entropy maps representing prediction uncertainty, as $W_{\text{off}}$ increases, entropy increases. All maps are obtained by averaging the softmax probabilities over \(N_{\rm seed}=10\) independently initialized CNNs and \(K=50\) disorder realizations for each parameter point.}
    \label{fig:rows2-6_results}
\end{figure}

\bibliography{apssamp}

\end{document}